\documentclass[preprint,12pt]{elsarticle}

 \usepackage{graphicx}
\usepackage{amssymb}
\usepackage{amsmath}
\usepackage{slashed}

\newcommand{\Li}[1]{\mathrm{Li}_2\left(#1\right)}
\newcommand{\Ln}[1]{\ln\left(#1\right)}
\newcommand{\leri}[1]{\left(#1\right)}
\newcommand{\Leri}[1]{\left[#1\right]}

\newcommand{\ord}[1]{\mathcal{O}\left(#1\right)}

\newcommand{\MS}[0]{$\overline{\mathrm{MS}}$}
\newcommand{\m}[0]{\overline{m}}
\newcommand{\x}[0]{\overline{x}}
\newcommand{\C}[0]{\mathtt C}
\journal{arXiv}
\begin{document}

\begin{frontmatter}



\title{The inclusive decay $b \to c \bar{c} s$ revisited}

\author[TUM]{Fabian Krinner}
\author[IPPP,CERN]{Alexander Lenz}
\author[TUM]{Thomas Rauh}

\address[TUM]{Physik-Department, Technische Universit\"at M\"unchen, James-Franck-Stra\ss e, 85748 Garching, Germany}
\address[IPPP]{Institute for Particle Physics and Phenomenology, Durham University, Durham DH1 3LE, UK}
\address[CERN]{CERN - Theory Division, PH-TH, Case C01600, CH-1211 Geneva 23}

\begin{abstract}
The inclusive decay rate $b \to c \bar{c} s$ is enhanced considerably 
due to perturbative QCD corrections. We recalculate the dominant part of the
NLO-QCD corrections, because they cannot be reconstructed from the 
literature and we give the full expressions in this paper. 
Further we include some previously neglected corrections originating from
penguin diagrams.
Combined with the impressive progress in the accurate determination of input
parameters like charm quark mass, bottom quark mass and CKM parameters,
this enables us to make a very precise prediction of the corresponding
branching ratio
$ \mathcal B ( b \to c\bar cs) = (23 \pm 2)\%$.
This result is an essential ingredient for a model and even decay channel 
independent search for new physics effects in B decays.

\end{abstract}


\end{frontmatter}

% \linenumbers

\section{Introduction}
\label{sec:intro}
Despite the impressive experimental achievements 
in flavour physics in recent years we still do not 
have compelling proof for new physics effects. 
Time by time interesting evidence for deviations has shown up
(see e.g. the status of new physics searches in $B$-mixing in
2010 \cite{Lenz:2010gu}). Unfortunately most of it
vanished as soon as more precise data became available
(see e.g. the status of new physics searches in $B$-mixing in
2012 \cite{Lenz:2012az}).
Currently the standard model 
\cite{Glashow:1961tr,Weinberg:1967tq,Salam:1968rm}
% REFERENCE FOR GLASHOW; WEINBERG; SALAM
and the CKM mechanism 
\cite{Cabibbo:1963yz,Kobayashi:1973fv} 
% REFERENCE FOR CABIBBO AND KOBAYASHi, MASAKAWA
seem to work at an unexpectedly precise level, see e.g. \cite{Eberhardt:2012gv,Baak:2012kk}
for standard model fits after the Higgs boson
\cite{Englert:1964et,Higgs:1964pj,Guralnik:1964eu}
% REFERENCE FOR ENGLERT, BROUT; HIGGS; GURALNIK, HAGEN, KIBBLE 
discovery
\cite{Aad:2012tfa,Chatrchyan:2012ufa}
% REFERENCEs FOR ATLAS AND CMS FOR THE HIGGS DISCOVERY
.
\\
Nevertheless most of the motivations for looking for an extension of
the standard model still remain valid, e.g. an explanation of the
baryon asymmetry in the universe or the nature of dark matter.
Flavour physics is a mandatory ingredient in the programme
of searching for and investigating extensions of the standard model.
Finding statistical significant deviations of measurements
from the corresponding standard model predictions
might provide the first evidence of new physics effects.
But even if new physics is detected in other fields, like
a direct production of new degrees of freedom at the LHC, then flavour physics
will be very helpful in pinning down the properties of the new particles.
Finally also in the worst case of finding no direct evidence of beyond standard 
model physics in the near future, the parameter space of hypothetical new physics models can be
shrunk considerably, some models might even be excluded, as it recently
happened for the case of a perturbative fourth generation of 
chiral fermions, see e.g.
\cite{Eberhardt:2012gv,Djouadi:2012ae,Kuflik:2012ai,Lenz:2013iha}.
One essential prerequisite for all this tasks is clearly the theoretical 
control of our standard model predictions.
Besides the already mentioned fact that there are no huge new physics
effects around the corner, we learnt from the LHC data a lot about our theoretical tools.
The applicability of the Heavy-Quark-Expansion (HQE) 
\cite{Shifman:1984wx,Khoze:1983yp,Khoze:1986fa,Chay:1990da,Bigi:1992su,Bigi:1993fe,Blok:1993va,Manohar:1993qn}
% REFERENCES FOR THE HQE
was questioned many times
in the literature, in particular for decays with a limited phase space in the final 
states, e.g. decays like $b \to c \bar{c} s$.
see e.g.  \cite{Falk:1994hm,Altarelli:1996gt,Dunietz:1996zn} .
Thus, an ideal testing ground for the HQE is the decay
rate difference $\Delta \Gamma_s$ of the neutral $B_s$ mesons. The dominant contribution
to this quantity is given by the decay channels $\bar{B}_s \to D_s^{(*)+} D_s^{(*)-}$
\cite{Aleksan:1993qp,Chua:2011er}
(triggered by the quark level decay $b \to c \bar{c} s$), with
an energy release of only about $1.4$ GeV. If the HQE does not converge, because
of a too large expansion parameter (1/energy release) then this should clearly show up
in $\Delta \Gamma_s$.
In 2012 LHCb made the first measurement (more than five standard deviations from zero) 
of the decay 
rate difference by investigating the decay $B_s \to J/ \psi \phi$
\cite{LHCb:2012xxx}, see \cite{LHCb:2011aa} for the previously published result. 
$\Delta \Gamma_s$  was also studied by the ATLAS Collaboration 
\cite{Aad:2012kba}\footnote{ATLAS presented a flavour tagged update of 
this result at BEAUTY 2013.},
the CDF Collaboration \cite{Aaltonen:2012ie}
and the D0 Collaboration \cite{Abazov:2011ry} and the values from LHCb were  
recently updated at Moriond 2013 \cite{LHCb:2013xxx}.
The Heavy Flavour Averaging Group (HFAG) \cite{Amhis:2012bh}
gives as an average the value
\begin{equation}
\Delta \Gamma_s^{Exp.} = + 0.081 \pm 0.011 \; \mbox{ps}^{-1} \; ,
\end{equation}
which excellently agrees with the standard model prediction
\cite{Lenz:2011ti}\footnote{This prediction is 
based on the NLO-QCD calculations in \cite{Lenz:2006hd,Beneke:2003az,Beneke:2002rj,Beneke:1998sy};
most of these results were confirmed in \cite{Ciuchini:2003ww}.}
\begin{equation}
\Delta \Gamma_s^{SM} = + 0.087 \pm 0.021 \; \mbox{ps}^{-1} \; .
\end{equation}
This clearly shows the validity of the HQE, in particular also for
the mistrusted decay channel $ b \to c \bar{c} s$, 
see \cite{Lenz:2012mb} for a more detailed discussion.
There are also some indications that the HQE might even work in
the charm sector  when applied to the lifetimes of $D$ mesons, see \cite{Lenz:2013aua}
for a very recent investigation.
\\
Another important aspect of the new experimental results is the fact, that
there is still plenty of room for NP effects.
In $B_s-$mixing we still could have new phases which are considerably larger than the 
standard model phases \cite{Lenz:2012az}. Also the investigation of rare decays
still leaves quite some sizable room for beyond standard model effects, see e.g.
\cite{Beaujean:2013qc,Altmannshofer:2012az,Descotes-Genon:2013hba} 
for some very recent constraints.
\\
To summarise the current situation (see also \cite{Bediaga:2012py} for a review): 
the desired huge new physics effects have not been
found, but there is still room for some sizable effects. In order to disentangle
such effects higher precision is mandatory both in experiment and theory. In that respect 
it is of course 
very promising that our theoretical tools have passed many non-trivial tests.

In this paper we provide some theoretical prerequisites for a model independent and even
decay channel independent search of new physics. We propose a re-investigation of inclusive
$b$-decays.
\\
Motivated by the experimental measurement of the dimuon asymmetry $A_{sl}^b$
from the D0 collaboration at Tevatron in 2010 \cite{Abazov:2010hv, Abazov:2010hj}
and 2011 \cite{Abazov:2011yk}, which differs by 3.9 standard deviations from the
standard model prediction given in \cite{Lenz:2006hd}\footnote{This prediction is 
based on the NLO-QCD calculations in \cite{Beneke:2003az,Beneke:2002rj,Beneke:1998sy};
these results were confirmed in \cite{Ciuchini:2003ww}.}
it became quite popular to investigate new physics models that enhance the absorptive
part of the $B_d$ and $B_s$ mixing amplitudes. Such an enhancement would also lead to 
considerable modifications of different $B$-meson decay channels.
One promising candidate was the decay $B_s \to \tau^+ \tau^-$, which was investigated e.g.
in \cite{Bobeth:2011st,Skiba:1992mg,Grossman:1996qj,Dighe:2010nj,Bauer:2010dga,Dighe:2012df}. 
Having stronger experimental bounds on this channel
would be very helpful, even if the significance of the deviation in the dimuon asymmetry
went down recently \cite{Borissov:2013wwa}.
\\
A large branching ratio for $B_s \to \tau^+ \tau^-$ would also affect several inclusive 
decay rates: 
the total decay rate $\Gamma_{tot} = 1/ \tau$, 
the charm-less  decay rate $\Gamma (b \to \mbox{no charm})$,
the semileptonic branching ratio $B_{sl} = \Gamma_{sl}/\Gamma_{tot}$ 
and the average number of charm quarks per $b$-decay $n_c$. 
Some time ago these quantities received quite some attention, see e.g.
\cite{Altarelli:1991dx,Bigi:1993fm,Falk:1994hm,Kagan:1994qg,Bagan:1994qw,Buchalla:1995kh,Neubert:1996we,
Dunietz:1996zn,Kagan:1997qna,Neubert:1998bm,Kagan:1997sg,Lenz:2000kv}
because there seemed to be some discrepancies between theory and experiment.
For the investigation of the inclusive branching ratios the NLO-QCD corrections turned out 
to be very important (see e.g. \cite{Voloshin:1994sn}), they were determined 
for $b \to c l^- \bar{\nu}$ analytically in \cite{Nir:1989rm},
for $b \to c \bar{u} d   $ in \cite{Bagan:1994zd},
for $b \to c \bar{c} s   $ in \cite{Bagan:1995yf},
for $b \to $ no charm      in \cite{Lenz:1997aa}
and 
for $b \to s g $           in \cite{Greub:2000sy,Greub:2000an}.
Since the above listed numerical analyses are at least 15 years old and there
was a lot of progress in the precise determination of the relevant standard model
parameters like CKM elements and quark masses an update of the numerical predictions
for inclusive  decay rates is clearly overdue. This case is strengthened by the 
experimental confirmation of the HQE.
Moreover an investigation of the inclusive branching ratios is not limited to a certain 
decay channel like $B_s \to \tau^+ \tau^-$, but it is sensitive to all possible decay rates, 
even if 
there would be some invisible decay channels.
Finally the most recent experimental number for $n_c$ stems from 
BaBar from 2006 using about $231 \cdot 10^6 \; B\bar{B}$ events \cite{Aubert:2006mp}.
Here clearly an experimental update is possible and also Belle as well as LHCb might 
investigate these quantities.
\\
Preparing a theoretical update we found, however, that the formulae given in the paper 
\cite{Bagan:1995yf} for the $b \to c \bar{c} s $ decay rate give IR divergent expressions. 
Thus we recalculate these important corrections in this work and include also previously 
neglected contributions. 
Comparing our numerical results (for the old input parameters) with the ones given 
in \cite{Bagan:1995yf}  we find an exact agreement. 
Hence, we conclude that there were simply several misprints in the formulae of 
\cite{Bagan:1995yf}. We give here the corrected expressions for 
$\Gamma (b \to c \bar{c} s)$
and finally we also perform the numerical analysis with up-to-date input parameters.
\\
The paper is structured as follows:
In Section \ref{sec:calc} we describe in detail the order $\ord{\alpha_s}$ 
calculation of the decay width of the channel $b\rightarrow c\bar cs$.
Besides discussing the diagrams that were already done in \cite{Bagan:1995yf},
we also calculated some previously neglected contributions
in Section \ref{sec:alphaPeng} and Section \ref{sec:alphaChrom}. Different quark mass schemes
are investigated in Section \ref{sec:MSbar}.
In Section \ref{sec:finalForm} we discuss our final result in detail. First we show 
explicitly the cancellation of the renormalisation scheme dependence in
Section \ref{sec:scheme}, next we compare our result in detail with \cite{Bagan:1995yf}
in Section \ref{sec:comparison} and point out the misprints in that work.
The up-to-date numerical results for the decay rate are presented and discussed 
in Section \ref{sec:numRes}.
Finally we conclude in Section \ref{sec:conclusion}.
Longer analytic expressions of our calculation are presented in the Appendix.

\section{Calculation of the inclusive $b\to c\bar cs$ decay width}
In this chapter we discuss the calculation of the inclusive
$b\to c\bar cs$ decay width up to order $\ord{\alpha_s}$ in detail, we
also include some previously neglected contributions.
\label{sec:calc}
\subsection{The effective Hamiltonian}
\label{sec:theory}
The starting point of our calculation is the effective weak Hamiltonian,
which can be written as (see e.g. \cite{buchallaburaslautenbacher}
for a review):
\begin{equation}
\mathcal{H}_\mathrm{eff}=\frac{G_F}{\sqrt{2}}\leri{\xi_c \sum_{i\in\{1,2\}} C_i Q_i -\xi_t \sum_{i\in \{3\ldots6,8\}}C_iQ_i}
,\end{equation} 
where $G_F$ denotes the Fermi constant, $\xi_q=V_{qb}V_{qs}^*$ represents the CKM-elements,
$C_i$ the Wilson coefficients  and $Q_i$ the appearing four quark operators.
The individual operators are given by:
\begin{equation}
\begin{array}[t]{r@{\hspace{1pt}}l}
Q_1&=\bar c_\alpha \Gamma_\mu b_\beta \otimes \bar s_\beta \Gamma^\mu	c_\alpha,\\
Q_2&=\bar c_\alpha \Gamma_\mu b_\alpha \otimes \bar s_\beta \Gamma^\mu	c_\beta	,\\
Q_3&=\bar s_\alpha \Gamma_\mu b_\alpha \otimes \bar c_\beta \Gamma^\mu	c_\beta	,\\
Q_4&=\bar s_\alpha \Gamma_\mu b_\beta \otimes \bar c_\beta \Gamma^\mu	c_\alpha,\\
Q_5&=\bar s_\alpha \Gamma_\mu b_\alpha \otimes \bar c_\beta \Gamma^\mu_+c_\beta	,\\
Q_6&=\bar s_\alpha \Gamma_\mu b_\beta \otimes \bar c_\beta \Gamma^\mu_+	c_\alpha,\\
Q_8&=-\displaystyle\frac{g_s}{8\pi^2}m_b \bar s_\alpha \sigma^{\mu\nu} \leri{1+\gamma_5} T_{\alpha\beta}^A b_\beta G_{\mu\nu}^A	
.\end{array}
\end{equation}
$\alpha$ and $\beta$ denote SU(3) colour indices and
$g_s$ is the strong coupling-constant. 
The appearing Dirac-structures are given by 
$\Gamma^\mu=\gamma^\mu\leri{1-\gamma_5}$, 
$\Gamma^\mu_+=\gamma^\mu\leri{1+\gamma_5}$ and 
$\sigma^{\mu\nu}=i/2 [\gamma^\mu,\gamma^\nu]$. 
\\
The Wilson coefficients can be expressed as a series in powers of $\alpha_s$:
\begin{equation}
C_i=C_i^{(0)}+\frac{\alpha_s}{4\pi}C_i^{(1)}+\ord{\alpha_s^2}
\label{eq:Wilson}
.\end{equation}
Note, that throughout this paper it is always understood, that the Wilson-coefficients 
are evaluated at the renormalisation-scale $\mu$, i.e.: $C_i=C_i\leri{\mu}$. 
The same holds for the strong coupling constant: $\alpha_s=\alpha_s\leri{\mu}$, if not stated otherwise.

  \subsection{The inclusive $b\to c\bar cs$ decay width}
  \label{sec:width}
  The inclusive decay width for the channel $b\to c\bar cs$ is given as
a phase space integration over the squared matrix element, describing the
transition of the $b$-quark into the final state $ c\bar cs$ via the weak 
Hamiltonian.
\begin{eqnarray}
\Gamma_{c\bar cs} & = &
\frac{8 \pi^4}{m_b} \int \prod \limits_{i=1}^3 
\left[ \frac{d^3p_i}{(2 \pi)^3 2 E_i}\right] 
\delta^{(4)} \left( p_B - \sum \limits_{i=1}^3 p_i \right)
\left| \langle  c\bar cs | {\cal H}_{eff} | b \rangle \right|^2
\, .
\end{eqnarray}
The above matrix element can be expanded in powers of the strong coupling
\begin{eqnarray}
{\cal M} & := & \langle  c\bar cs | {\cal H}_{eff} | b \rangle
\nonumber
\\
         & = & {\cal M}^{(0)} 
             + \frac{\alpha_s}{4 \pi} {\cal M}^{(1)} 
             + {\cal O} \left(\frac{\alpha_s}{4 \pi} \right)^2 \; ,
\end{eqnarray}
thus we get for the squared matrix element
\begin{equation}
\left| {\cal M} \right|^2 = {\cal M}^{(0)\dagger} {\cal M}^{(0)} 
+  \frac{\alpha_s}{4 \pi}
\left( {\cal M}^{(0)\dagger} {\cal M}^{(1)} +
       {\cal M}^{(1)\dagger} {\cal M}^{(0)} \right)
+ {\cal O} \left(\frac{\alpha_s}{4 \pi} \right)^2 \; .
\label{Eq:expand}
\end{equation}
Therefore the inclusive decay width for the channel 
$b\to c\bar cs$ can be written as:
\begin{equation}
\Gamma_{c\bar cs} = \Gamma^{(0)}_{c\bar cs} + \frac{\alpha_s}{4\pi} \Gamma^{(1)}_{c\bar cs}+\ord{\alpha_s^2}
,
\end{equation}
with the leading-order (LO) contribution $\Gamma^{(0)}_{c\bar cs}$ and the 
sizable next-to-leading-order (NLO) correction  $\Gamma^{(1)}_{c\bar cs}$.
The LO decay width is given as:
\begin{equation}
\Gamma_{c\bar cs}^{(0)}= \Gamma_0|\xi_c|^2g\mathcal{N}_a%%%CITE
.\end{equation}
The factor $\Gamma_0$ is the width of a decay into three mass- 
and colourless particles:
\begin{equation}
\Gamma_0=\frac{G_F^2m_b^5}{192\pi^3}
.
\end{equation}
The factor $\mathcal{N}_a$ stems from the leading term in Eq.(\ref{Eq:expand}),
it is a linear combination of products of LO Wilson coefficients weighted with 
colour factors.
We include here only the contribution of the tree level operators
$Q_1$ and $Q_2$, the penguin operators will be treated as a QCD correction
and discussed below.
Thus our result agrees, up to the phase space function, with the 
result for the decay $b \to c \bar{u} d$, see e.g. \cite{Bagan:1994zd}:
\begin{equation}
\mathcal{N}_a=3C_1^{(0)2}+3C_2^{(0)2}+2C_1^{(0)}C_2^{(0)}
.
\end{equation}
The function $g$ is the tree-level-phase-space-integral depending on the final-state-masses. Neglecting the strange quark mass, $g$ reads
(see e.g. Eq.(4.2) of \cite{Bagan:1994zd} or \cite{Gourdin:1979qq} for an 
early reference):
\begin{equation}\begin{array}[t]{c}
g=\sqrt{1-4x_c^2}\leri{1-14x_c^2-2x_c^4-12x_c^6}
\\%%%CITE
+24 x_c^4\leri{1-x_c^4}
\Ln{\displaystyle\frac{1+\sqrt{1-4x_c^2}}{1-\sqrt{1-4x_c^2}}}
,\end{array}
\label{Eq:phase_LO}
\end{equation}
where $x_c=m_c/m_b$ is the ratio of the charm- and the bottom-quark mass.
\\
The NLO correction can be split up into several contributions:
\begin{equation}\label{eq:gamma1}
\Gamma^{(1)}_{c\bar cs}=\Gamma_{c\bar cs}^{\alpha_s} + \overline{\Gamma}_{c\bar cs}^m + \Gamma_{c\bar cs}^\mathrm{PO} + \Gamma_{c\bar cs}^{C} + \Gamma_{c\bar cs}^\mathrm{PI} + \Gamma_{c\bar cs}^{Q_8}
.\end{equation}
The different terms of Eq.(\ref{eq:gamma1}) are sorted according to their size.
$\Gamma_{c\bar cs}^{\alpha_s}$ denotes corrections from the one gluon 
exchange in the LO diagrams within the effective theory. 
$\overline{\Gamma}_{c\bar cs}^m$ describes corrections of 
order $\ord{\alpha_s}$ stemming from the translation of the $b$-quark mass
from the pole- into the \MS-scheme \cite{Bardeen:1978yd}. 
This term vanishes by definition if the 
pole-scheme is used for the $b$-quark mass. 
$\Gamma_{c\bar cs}^\mathrm{PO}$ contains LO contributions from 
the penguin operators $Q_{3\ldots6}$, 
which are treated as $\ord{\alpha_s}$-effects in this paper. 
$\Gamma_{c\bar cs}^C$ describes NLO corrections to the 
Wilson-coefficients $C_{1,2}$. 
Besides these contributions, which were already calculated in
\cite{Bagan:1995yf}
we also determined two previously neglected  corrections:
$\Gamma_{c\bar cs}^\mathrm{PI}$ and $\Gamma_{c\bar cs}^{Q_8}$ stem 
from insertions of the operators $Q_{1,2}$ and the operator $Q_8$ 
in penguin diagrams within the effective theory.
\\
Below all these contributions will be discussed in detail.

  \subsection{Gluon-corrections to the insertion of $Q_{1,2}$ into treelevel diagrams}
  \label{sec:alphaTree}
  The largest $\ord{\alpha_s}$-contribution stems directly from the 
next-to-leading term in Eq.(\ref{Eq:expand}). It arises
from one-gluon corrections 
to the insertion of the operators $Q_1$ and $Q_2$ in tree-level diagrams
of the effective theory, see Fig.(\ref{pic:analog}). 
Hence, this term can be written as:
\begin{equation}
\begin{array}[t]{c}
\Gamma_{c\bar cs}^{\alpha_s} = 8 \Gamma_0 |\xi_c|^2 
\cdot\leri{C_2^{(0)2}g_{22}+C_1^{(0)2}g_{11} + \frac{2}{3}C_1^{(0)}C_2^{(0)}g_{12}}
.
\end{array}
\label{Eq:alphaTree}
\end{equation}
\begin{figure}[h]
  \centering
  \fbox{
    \includegraphics[width=0.6\textwidth]{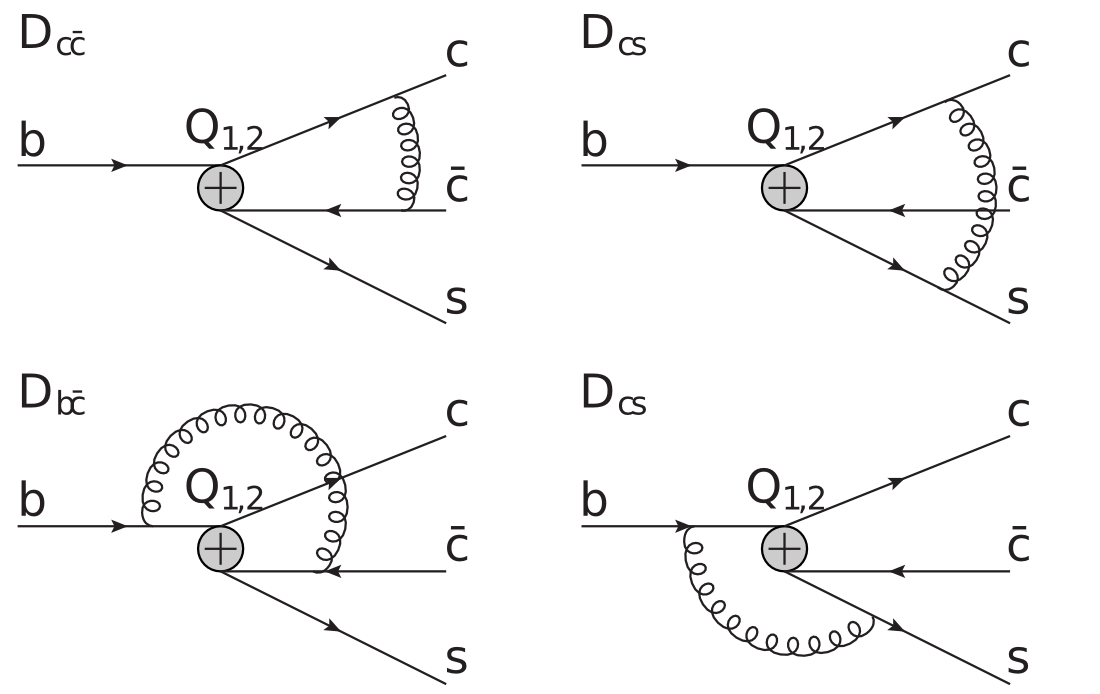}
  }
  \caption{Virtual gluon corrections to the insertion of the operators
$Q_1$ and $Q_2$ in tree-level diagrams of the effective theory.
Only the diagrams that could not be taken from \cite{hokimpham1984}
and had to be be calculated anew, are shown. 
The corresponding diagrams with real gluon-emission are also not pictured.}
\label{pic:analog}
\end{figure}
Since we strictly expand in $\alpha_s$ and discard terms of 
${\cal O}(\alpha_s^2)$, we use in Eq.(\ref{Eq:alphaTree}) 
the LO expressions for the Wilson 
coefficients, denoted by $C_1^{(0)}$ and $C_2^{(0)}$.
The functions $g_{11}$, $g_{12}$ and $g_{22}$ are given by 
phase-space-integrals over the corresponding diagrams, 
where one gluon is exchanged.
For massless quarks these corrections were already calculated in 1981
by Altarelli et al. \cite{Altarelli:1980fi} and later confirmed in
\cite{Buchalla:1992gc},
\begin{eqnarray}
g_{11} & = & \frac{31}{4} - \pi^2 = g_{22} \; ,
\\
g_{12} & = & -\frac{7}{4} - \pi^2 \; .
\end{eqnarray}
For massive quarks the calculation is much more complicated and it was
performed for the first time in 1995 \cite{Bagan:1995yf}.
However, most of the required corrections can be inferred from the work 
\cite{hokimpham1984} by \textit{Hokim} an \textit{Pham}. 
There the exact width for a weak decay into three particles with arbitrary 
masses has been calculated up to order $\ord{\alpha_s}$ in the full 
standard model, ignoring, however, the momentum-square in the 
$W^\pm$-propagator (as in the effective theory).
Thus their result corresponds to the double insertion of the operator $Q_2$
into the next-to-leading term in Eq.(\ref{Eq:expand}). 
Using the notation of \cite{hokimpham1984} the factor $g_{22}$ can be 
expressed as
\begin{equation}
\label{eq:HPinput}
g_{22} = \frac{4}{\Omega_0} \leri{\Gamma_l+\Gamma_u}
.
\end{equation}
The contributions on the right hand side of Eq.(\ref{eq:HPinput}) are given
in the formul\ae\ (3.9), (3.36) and (4.34) of \cite{hokimpham1984}. 
The arguments of theses functions have to be chosen as:
\begin{equation}\label{eq:rho123}
\rho_1=\leri{\frac{m_c}{m_b}}^2,\ 
\rho_2=0,\ 
\rho_3=\leri{\frac{m_c}{m_b}}^2
.
\end{equation}
Further information about using the results from \cite{hokimpham1984}
for the calculation of $\Gamma_{c \bar{c} s}$  is 
given in \ref{sec:pedestrian}.
\\
Also the contributions from double insertions of the operator $Q_1$ 
can be inferred from \cite{hokimpham1984}, if the diagrams are
Fierz-transformed.
To do so, it was crucial to choose the evanescent operators in 
such a  way that Fierz-symmetry is maintained at the one-loop-level.
Following \cite{burasweisz1990} we use:
\begin{equation}
\label{eq:evan}
E= \gamma_\mu\gamma_\nu\Gamma_\sigma\otimes\gamma^\sigma\gamma^\nu\Gamma^\mu-\leri{4-8\varepsilon}\Gamma_\mu\otimes\Gamma^\mu
,\end{equation}
where $\varepsilon$ is the usual regulator in dimensional 
regularisation in $D=4-2\varepsilon$ dimensions.
Then we get for $g_{11}$:
\begin{equation}\label{eq:HPinput'}
g_{11} = \frac{4}{\Omega_0} \leri{\Gamma'_l+\Gamma'_u}
,\end{equation}
The contributions on the right hand side of Eq.(\ref{eq:HPinput'}) are also
given in the formul\ae\ (3.9), (3.36) and (4.34) of \cite{hokimpham1984},
but  now the arguments differ from (\ref{eq:rho123}), which is denoted 
by the apostrophe. 
They read instead:
\begin{equation}\label{eq:rho123'}
\rho'_1=0,\ \rho'_2=\leri{\frac{m_c}{m_b}}^2,\ \rho'_3=\leri{\frac{m_c}{m_b}}^2
.
\end{equation}
Note, that in \cite{hokimpham1984} diagrams  containing a gluon exchange 
between the two different fermion-lines (i.e. the $b \to c$-line and 
the $\bar{c} \to s$-line) vanished due to the colour factor . 
As already mentioned, the results in \cite{hokimpham1984} directly
correspond to a double insertion of the operator $Q_2$ and they can
also be used for the double insertion of the operator $Q_1$.
However, in our calculation also a mixed contribution, i.e. an insertion 
of $Q_1$ and $Q_2$, arises, where the colour structure does not vanish,
if a gluon is exchanged between the two different fermion-lines. 
The part of the contributions to $g_{12}$, where the gluon does not connect
the two fermion lines can be extracted again from \cite{hokimpham1984}.
Four diagrams, where the gluon couples to both of the 
appearing fermion-lines,  have to be computed new.
These are the diagrams pictured in Figure \ref{pic:analog}. 
The four new contributions are denoted by
$g_{c\bar c}$, $g_{cs}$, $g_{b\bar c}$ and $g_{bs}$ 
and they are  explicitly given in the \ref{sec:radAn}. 
Hence, we get for the last contribution in Eq.(\ref{Eq:alphaTree})
\begin{equation}
g_{12} = g_{22} + g_{c\bar c} + g_{cs} + g_{b\bar c} + g_{bs}
.\end{equation}
These contributions have been calculated for the decay $b \to c \bar{u} d$ in 
\cite{Bagan:1994zd}. Here one particle in the final state had to be taken
massive. The  decay $b \to c \bar{c} s$ was investigated in 
\cite{Bagan:1995yf}, where all particles in the final state are massive.
Unfortunately the results in \cite{Bagan:1995yf} contain several misprints
and cannot be used for a numerical reanalysis, so they had to be calculated 
anew.

  \subsection{The $b-$quark mass in the $\overline{\mathrm{MS}}$-scheme}
  \label{sec:MSbar}
  Since the pole mass suffers from renormalon ambiguities, as discussed 
e.g. in \cite{Beneke:1994sw,Bigi:1994em}, it has been argued, that the pole mass 
scheme is not well suited for performing precise calculations of inclusive
decay rates, while 
short distance masses like the \MS-mass \cite{Bardeen:1978yd} are better 
suited for this purpose \cite{Bigi:1994em}. 
The relation between the pole mass scheme and the \MS-mass scheme is given as 
a power series in the strong coupling. 
Up to the order $\ord{\alpha_s}$ it reads
\begin{equation}
\label{eq:mscheme}
\begin{array}[t]{c}
m_q^\mathrm{pole}=\m_q\leri{\mu_m}
\leri{1+\displaystyle\frac{\alpha_s\leri{\mu_m}}{\pi}
\cdot\Leri{\displaystyle\frac{4}{3}-\Ln{\displaystyle\frac{\m_q\leri{\mu_m}^2}{\mu_m^2}}}}
.
\end{array}
\end{equation}
Therefore the translation between the two 
schemes creates additional $\ord{\alpha_s}$-corrections in the total inclusive decay rates.
These terms can be written as:
\begin{equation}\begin{array}[t]{c}
\overline\Gamma_{c\bar cs}^m =
\Gamma_{c\bar cs}^{(0)}\Big[\displaystyle\frac{80}{3} - 20 \Ln{\frac{m_b^2}{\mu_m^2}}
-8x_c\Ln{x_c}\displaystyle\frac{d\Ln{g}}{dx_c}\Big]
\; ,
\end{array}
\label{eq:masscorr}
\end{equation}
where $g$ is the phase space function from Eq.(\ref{Eq:phase_LO}).
In LO there is a strong dependence on the scale $\mu_m$, that appears
in Eq.(\ref{eq:mscheme}).
The relation between the two mass schemes is, however, known very precisely -
corrections up to order $\ord{\alpha_s^3}$ have been calculated in \cite{MSrelation},
resulting in a small remaining scale dependence.
Hence we consider the large LO scale dependence in Eq.(\ref{eq:mscheme}) to be artificial
and we do not vary that scale for our final numerics, but we set $\mu_m=m_b$ during the calculation.
Thus the large logarithm in Eq. (\ref{eq:masscorr}) vanishes.
\\
Besides these two  mass-schemes, we have also investigated three other quark mass 
schemes, 
the kinetic                   \cite{Bigi:1996si},
the potential-subtracted (PS) \cite{PSMass} 
and the $\Upsilon(1S)$-scheme \cite{1SMass}
for our calculation, see Section \ref{sec:numRes}.

  \subsection{Contributions from the penguin-operators $Q_{3\ldots6}$}
  \label{sec:Q36}
  Besides the already discussed tree level diagrams, the
decays $b \to c \bar{c} s$ can also occur via penguin
diagrams. Therefore further contributions to the $b\rightarrow c\bar cs$ 
width arise 
from the insertion of QCD penguin-operators $Q_{3\ldots 6}$ into
tree diagrams of the effective theory. They can be written as:
\begin{equation}
\Gamma_{c\bar cs}^\mathrm{PO}=\displaystyle\frac{4\pi}{\alpha_s}\Gamma_0 
\Leri{\leri{\Re\leri{\xi_c\xi_t^*}\mathcal{N}_b + |\xi_t|^2\mathcal{N}_c} g + \leri{\Re\leri{\xi_c\xi_t^*}\mathcal{N}_d + |\xi_t|^2\mathcal{N}_e} g_+}
\label{Eq:peng}
.\end{equation}
Since this contribution is numerically of the order
of the $\alpha_s$-corrections, we decided to treat it formally as a 
QCD correction.
But as  no explicit factor $\alpha_s/4\pi$ appears in the calculation
we need the factor $4\pi/\alpha_s$ in Eq.(\ref{Eq:peng}) to 
cancel the corresponding artificial factor in
Eq.(\ref{eq:gamma1}).
The terms $\mathcal{N}_{b\ldots e}$ are again combinations of 
Wilson coefficients and colour factors 
\begin{eqnarray}
\label{eq:NN+1}
\mathcal N_b &=&-2\Big(3C_1C_3 + C_1C_4 + 3 C_2C_4+C_2C_3\Big),
\\
\label{eq:NN+2}
\mathcal N_c&=& \, \, \, \, \, \, \, \, \, \, \, 
   3C_3^2 +3C_4^2+2 C_3 C_4+ 3C_5^2 +3C_6^2 +2 C_5C_6 ,
\\
\label{eq:NN+3}
\mathcal N_d&=& -2 \Big(3 C_1C_5 +C_1C_6+3C_2C_6+C_2C_5\Big),
\\
\label{eq:NN+4}
\mathcal N_e&=& \, \, \, \, \,2 \Big(3 C_3C_5 + C_3C_6 + 3 C_4C_6 + C_4C_5\Big)
.\end{eqnarray}
Here formally terms of order $\ord{\alpha_s^2}$ appear, stemming from 
expressions of the form $C_i \cdot C_j$. These higher order terms have been 
discarded in the calculation to ensure a strict expansion up to  
order $\ord{\alpha_s}$. They were only kept in 
Eq.(\ref{eq:NN+1}-\ref{eq:NN+4}) to enable 
a more compact notation. The appearing Wilson-coefficients $C_{3\ldots 6}$, 
as well as the coefficient $C_8^{(0)}$ appearing in Section 
\ref{sec:alphaChrom}, can be taken from \cite{buchallaburaslautenbacher}.
\\
The function $g$ is the usual phase space 
factor, appearing first in Eq.(\ref{Eq:phase_LO}), while
$g_+$ is a different phase space integral, that appears when, 
an odd number of the factor $\leri{1+\gamma_5}$ appears in the 
Dirac-structure. 
Its analytic expression, which was to our knowledge not shown before, reads:
\begin{equation}\begin{array}[t]{c}
g_+ =4 x_c^2\bigg[\sqrt{1-4x_c^2}\leri{1+5x_c^2-6x_c^4} \\
+ 12 x_c^2 \leri{1-2x_c^2 +2 x_c^4}\Ln{\displaystyle\frac{2x_c}{1+\sqrt{1-4x_c^2}}}\bigg]
.\end{array}\end{equation}
Numerically $g_+$ agrees with Eq.(12) of \cite{Bagan:1994qw}.
Finally in Eq.(\ref{Eq:peng}) no approximation in the CKM-structure was used, 
as it is often done in the literature to eliminate $\xi_t$:
\begin{equation}
\xi_t = -\xi_c-\xi_u \approx -\xi_c
.\end{equation}
Using this approximation our results in 
Eqs.(\ref{eq:NN+1}-\ref{eq:NN+4}) 
agree with the ones 
in the erratum of \cite{Bagan:1994qw} and Eq.(\ref{eq:NN+1}) 
and Eq.(\ref{eq:NN+3})
agree with Eq.(XVII.14) from \cite{buchallaburaslautenbacher}.
  \subsection{Order $\ord{\alpha_s}$ corrections to the Wilson coefficients $C_{1,2}$}
  \label{sec:alphaWilson}
  Since the Wilson coefficients $C_{1}$ and $C_2$ are given as a series in the 
strong coupling, see Eq.(\ref{eq:Wilson}), 
they give rise to additional corrections of the order $\ord{\alpha_s}$. 
These corrections are given by:
\begin{equation}\label{eq:GammaC}\begin{array}[t]{c}
\Gamma_{c\bar cs}^C = \Gamma_0 |\xi_c|^2 g 
\Big[\Big(\displaystyle\frac{\alpha_s\leri{M_W}}{\alpha_s\leri{\mu}}-1\Big)
\cdot\Big(4C_+^{(0)2}R_++2C_-^{(0)2}R_-\Big)\\
+\Big(4C_+^{(0)2}B_+ + 2C_-^{(0)2}B_-\Big)\Big]
.
\end{array}
\end{equation}
These contributions arise in all NLO QCD corrections to non-leptonic
inclusive decays that are triggered via
tree-level diagrams (with appropriate phase space functions), e.g. 
$b \to c \bar{u} d$ \cite{Bagan:1994zd}
and
$b \to c \bar{c} s$ \cite{Bagan:1995yf}.
To simplify the notation we use here linear combinations of the operators 
$Q_1$ and $Q_2$, that do not mix under renormalisation, see e.g.
\cite{burasweisz1990}. This leads to the coefficients $C_+$ and $C_-$, 
which are defined as:
\begin{equation}
C_\pm = C_2\pm C_1
.
\end{equation}
These coefficients can be written as (see e.g. \cite{Bagan:1994qw})
\begin{equation}
C_\pm=C_\pm^{(0)} 
\Bigg(1+\frac{\alpha_s\leri{M_W}-\alpha_s\leri{\mu}}{4\pi}R_\pm+ 
\frac{\alpha_s\leri{\mu}}{4\pi}B_\pm\Bigg)
,
\end{equation}
with the leading order Wilson coefficients $C_\pm^{(0)}$, given by:
\begin{equation}
C_\pm^{(0)}=
\leri{
\frac{\alpha_s\leri{M_W}}{\alpha_s\leri{\mu}}}^{\gamma_\pm^{(0)}/(2\beta_0)}
.\end{equation}
$\beta_0$ is the leading coefficient of the QCD beta-function
and $ \gamma_\pm^{(0)}$ is the leading term of the anomalous dimensions
of the operators $Q_+$ and $Q_-$.
The coefficients read: 
\begin{equation}
\beta_0=11-\frac{2}{3}n_f=\frac{23}{3}
,\end{equation}
\begin{equation}
\gamma_+^{(0)}=4,\ \gamma_-^{(0)}=-8
,\end{equation}
\begin{equation}
R_+=\frac{10863-1278n_f+80n_f^2}{6(33-2n_f)^2}=\frac{6473}{3174}
,\end{equation}
\begin{equation}
R_-=-\frac{15021-1530n_f+80n_f^2}{3(33-2n_f)^2}=-\frac{9371}{1587}
,\end{equation}
\begin{equation}\label{eq:B}
B_\pm=\pm B\frac{N_C\mp1}{2N_C}
,\end{equation}
where $n_f=5$ is the number of active quark flavours and $N_C=3$ 
is the number of colours. 
The coefficient $B$ from Eq.(\ref{eq:B}) is the only 
scheme dependent quantity in Eq.(\ref{eq:GammaC}). 
In the NDR-scheme with the choice of evanescent operators as in 
Eq.(\ref{eq:evan}) the scheme dependent factor $B$ reads
$B^\mathrm{NDR}=11$.

  \subsection{Insertions of $Q_{1,2}$ into penguin diagrams}
  \label{sec:alphaPeng}
  Besides the corrections, in which the tree level diagrams are simply 
dressed with additional gluon lines 
(as discussed in Section \ref{sec:alphaTree}), 
new contributions arise at 
order $\ord{\alpha_s}$ due to insertions of the operators $Q_{1,2}$
into penguin diagrams, see Fig. {\ref{pic:penguin}.
\begin{figure}[ht]
\begin{center}
\fbox{
\begin{minipage}[b]{0.45\linewidth}
\centering
\includegraphics[width=\textwidth]{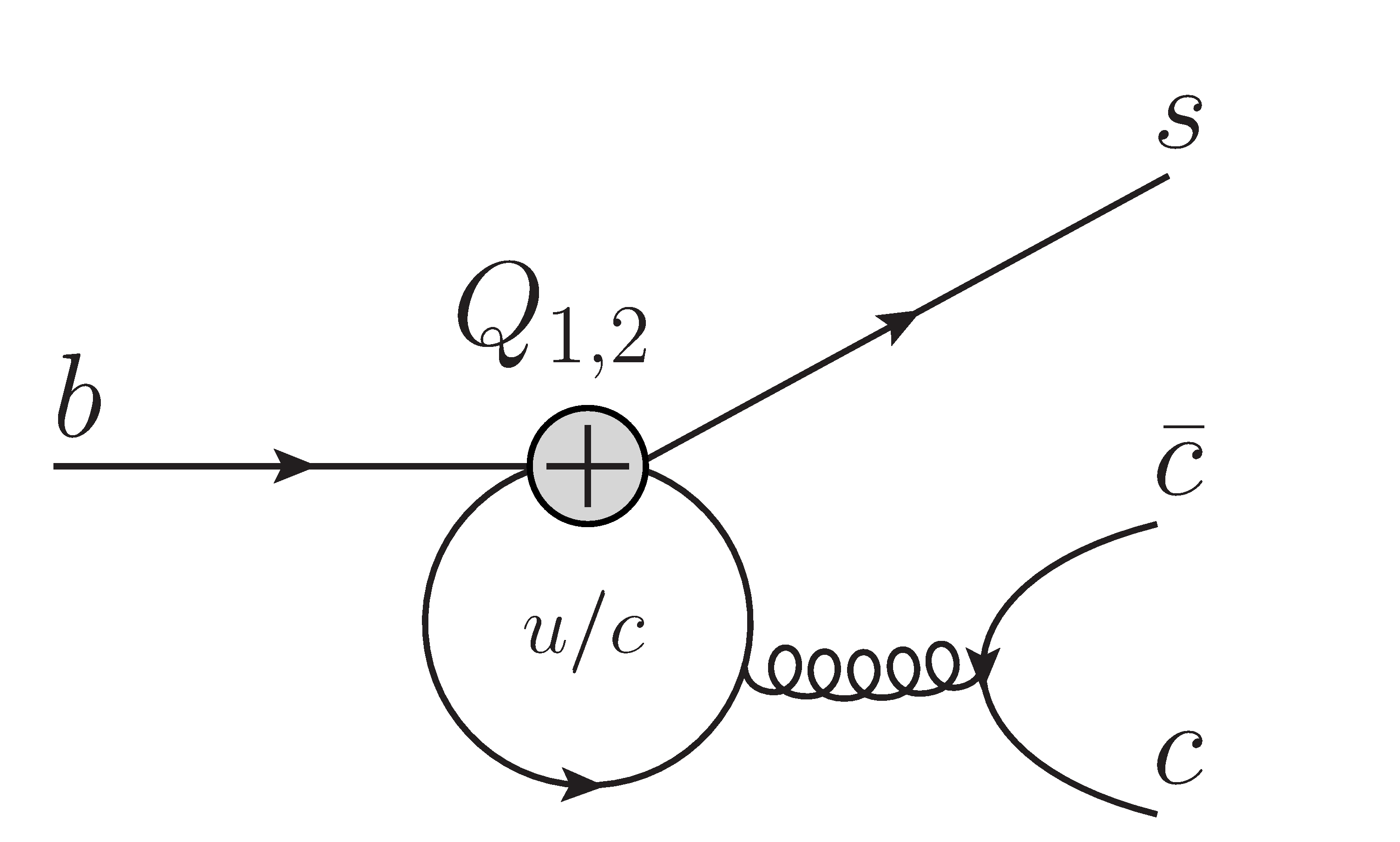}
\caption{Insertion of the operators $Q_1$ and $Q_2$ into a penguin
           diagram of the effective theory.
           Compared to the penguin operators $Q_3$ - $Q_6$ these 
           contributions appear with the large Wilson coefficient 
           $C_2$.}
\label{pic:penguin}
\end{minipage}
\hspace{0.5cm}
\hfill
\begin{minipage}[b]{0.45\linewidth}
\centering
\includegraphics[width=\textwidth]{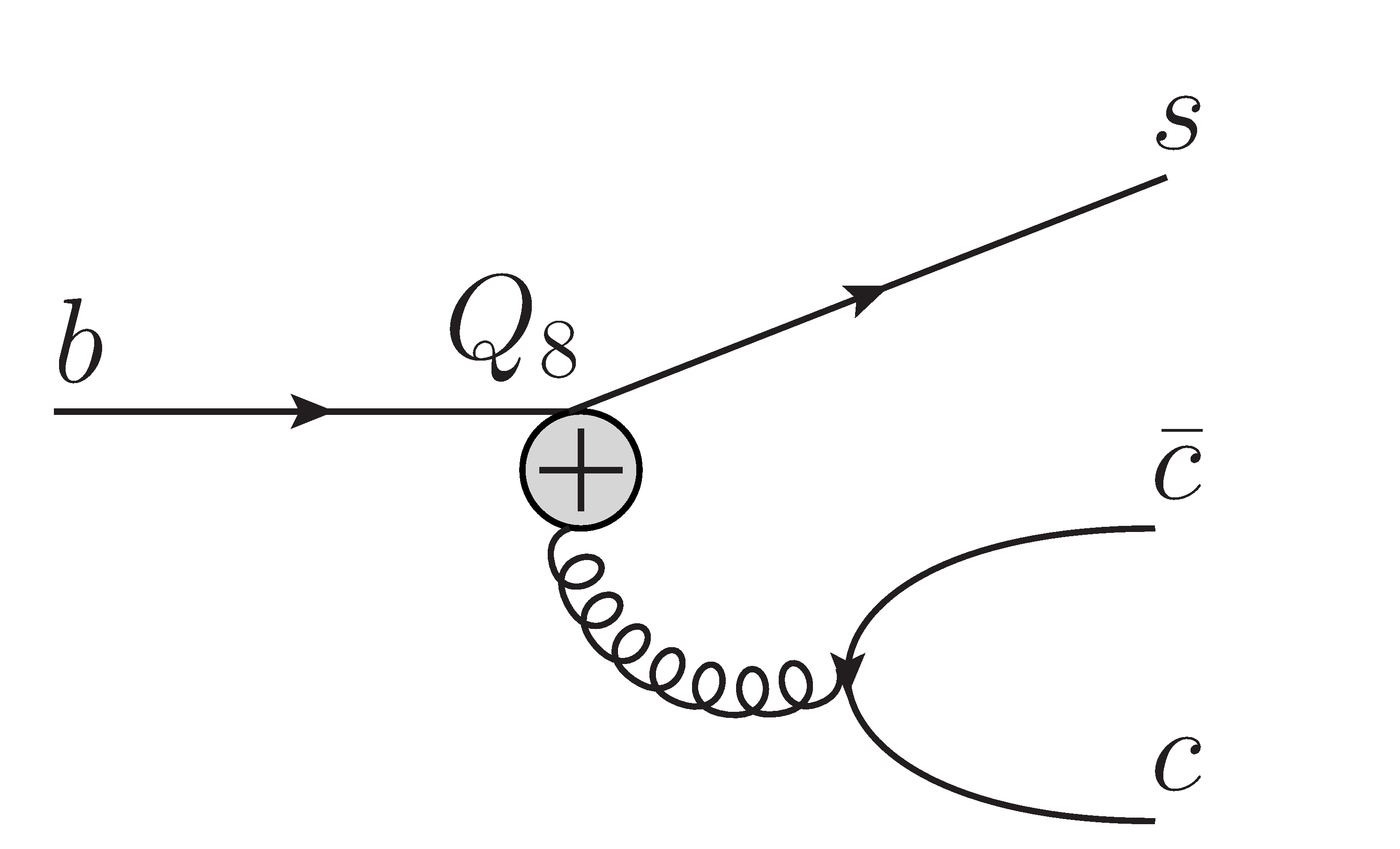}
\caption{Insertion of the chromomagnetic operator $Q_8$ into a 
           tree
           diagram of the effective theory. \vspace{1.25cm}
           }
\label{pic:q8}
\end{minipage}
}
\end{center}
\end{figure}
In principle both up and charm quarks could run in the penguin loop, 
but because of the strong CKM-suppression, only the case with a 
charm quark is considered here. 
Insertions of this kind vanish for $Q_1$, 
since they involve a vanishing trace of a generator of the colour SU(3), 
so only $Q_2$ has to be considered.
\\
The total contribution due to the insertion in penguin diagrams 
can then be written as:
\begin{equation}
\Gamma_{c\bar cs}^\mathrm{PI}=\frac{16}{3} \Gamma_0|\xi_c|^2
 \Re\Leri{C_2^{(0)2}\leri{g_{\mathrm{PI}}+g_{\mathrm{PI}+}}}
,\end{equation}
where the functions $g_{\mathrm{PI}}$ and $g_{\mathrm{PI}+}$ 
are two phase-space-functions, which were calculated analytically 
for the first time. The exact expressions are given in \ref{sec:pengAn}.
\\
This corrections have until now only been included in the calculation
of charm-less inclusive decay rates in \cite{Lenz:1997aa},
where they can be the dominant effect. Since the insertions into
penguin diagrams are also expected to give sizable contributions
to the $b$-quark decay into two charm quarks, we present them here 
for the first time.
In our calculation the same loop integrals arise as in \cite{Lenz:1997aa},
but the phase space integration has to be performed for massive
final state particles, while in \cite{Lenz:1997aa} only massless
final states were investigated. 
The insertion of the operator $Q_2$ in a penguin diagram
can be decomposed in the following way:
\begin{equation}
Q_P= \sum_{i=3}^6 r_{i}\leri{s_{c\bar c}} Q_i
.\end{equation}
$Q_P$ denotes the operator generated by the penguin insertion and 
$r_i\leri{s_{c\bar c}}$ are the appearing loop-functions which depend 
on the centre of mass energy $s_{c\bar c}$ of the $c$-$\bar c$-system. 
For the different $r_i$ the following relation holds:
\begin{equation}
r_4=r_6=-3r_3=-3r_5
.\end{equation}
The function $r_4$ was calculated in \cite{Lenz:1997aa} as
\begin{equation}\begin{array}[t]{c}
r_{4} = \displaystyle\frac{1}{9 s_{c\bar c}^{3/2}} \bigg[-3 \sqrt{s_{c\bar c}-4m_c^2}\leri{2m_c^2+s_{c\bar c}}
\Ln{\displaystyle\frac{\sqrt{1-4\frac{m_c^2}{s_{c\bar c}}}-1}{\sqrt{1-4\frac{m_c^2}{s_{c\bar c}}}+1}}\\
+\sqrt{s_{c\bar c}}\bigg(-2\leri{6m_c^2+s_{c\bar c}}+3s_{c\bar c}\Ln{\displaystyle\frac{m_c^2}{\mu^2}}\bigg) \bigg]
.\end{array}\end{equation}
As in the case of the LO contributions discussed in Section \ref{sec:Q36}
two different functions appear for the insertions into penguin diagrams, 
depending on whether the number of vertices that involve the 
factor $(1+\gamma_5)$ is even ($g_{\mathrm{PI}}$) 
or odd ($g_{\mathrm{PI}+}$).

  \subsection{Contributions from the chromomagnetic operator $Q_8$}
  \label{sec:alphaChrom}
  Additional order $\ord{\alpha_s}$ contributions arise through 
insertions of the chromomagnetic operator $Q_8$, see Figure \ref{pic:q8}.
After a straight-forward calculation, these contributions can be written as:
\begin{equation}\begin{array}[t]{c}
\Gamma_{c\bar cs}^{Q_8}=3 \Gamma_0 C_8^{(0)}g_{Q8}
\Leri{\Re\leri{\xi_t\xi_c^*}C_2^{(0)}-|\xi_t|^2\leri{C_4^{(0)}+C_6^{(0)}}}
.\end{array}\end{equation}
The phase space factor $g_{Q8}$, which was calculated here for the first time, 
is given as:
\begin{equation}\begin{array}[t]{c}
g_{Q8} =32 
\Leri{\displaystyle\frac{1 - 20 x_c^2 + 52 x_c^4 + 48 x_c^6}{9\sqrt{1-4x_c^2}}+ \displaystyle\frac{8}{3}x_c^4\leri{2 x_c^2-3}\Ln{\frac{2x_c}{1+\sqrt{1-4x_c^2}}}}
.\end{array}\end{equation}

\section{Investigation of the NLO-result for $\Gamma_{c\bar cs}$}
With all the contributions, which were calculated in 
Section \ref{sec:calc}, 
the total decay width of the inclusive decay 
$b\rightarrow c\bar cs$ can be written in NLO-QCD as:
\begin{equation}
\Gamma_{c\bar cs}=\Gamma_{c\bar cs}^{(0)}+\frac{\alpha_s}{4\pi}\leri{\Gamma_{c\bar cs}^{\alpha_s}+\overline{\Gamma}_{c\bar cs}^m+\Gamma_{c\bar cs}^\mathrm{PO}+\Gamma_{c\bar cs}^{C}+\Gamma_{c\bar cs}^\mathrm{PI}+\Gamma_{c\bar cs}^{Q_8}}
.\end{equation}
In the following chapter, the obtained results will be investigated in detail.
First we provide some consistency checks, 
next we perform a numerical analysis of the inclusive decay rate
and we also discuss several conceptual
issues, like different definitions of quark masses.

\label{sec:finalForm}
  \subsection{Scheme-independence}
%  \label{sec:checks}
     \label{sec:scheme}
     Here we show the independence of our results on the renormalisation scheme 
used for $\gamma_5$. 
The calculation of the correction $g_{12}$ in Section \ref{sec:alphaTree} 
involved UV-divergences and the result depends on the treatment of $\gamma_5$ 
in dimensional regularisation. The only source of these divergences
is the loop integral $\C_\mathtt{00}$ in the notation of
{\it LoopTools} \cite{hahnperez1999}.
Thus, the scheme dependence of the overall result is completely encoded in the terms:
\begin{equation}
\C_\mathtt{00}+B_1
,\end{equation}
for $g_{c\bar c}$ and $g_{bs}$ and
\begin{equation}
\C_\mathtt{00}+B_2
,\end{equation}
for $g_{cs}$ and $g_{b\bar c}$. The terms $B_1$ and $B_2$ contain all the scheme-dependence. 
Since they are just numbers, the phase space integration can be performed analytically, 
and one obtains for the total scheme dependent part:
\begin{equation}
\Gamma_{c\bar cs}^{\alpha_s\mathrm{scheme}}=
\frac{16}{3} \Gamma_0 |\xi_c|^2 g \leri{8B_1-32B_2}C_1^{(0)}C_2^{(0)}
.\end{equation}
This scheme dependence has to be cancelled by the scheme dependence of the 
Wilson coefficients $C_{1,2}$, which can be obtained from (\ref{eq:GammaC}) to be:
\begin{equation}
\Gamma_{c\bar cs}^{C\mathrm{scheme}}=\frac{16}{3} \Gamma_0|\xi_c|^2 g B C_1^{(0)}C_2^{(0)}
.\end{equation}
For the whole result to be scheme-independent, the following combination needs 
to have the same value in all schemes:
\begin{equation}
B+8B_1-32B_2
.\end{equation}
In the NDR and in the t'Hooft-Veltman-scheme one gets \cite{burasweisz1990}
\begin{equation}
\begin{array}{lcrlcrlcc}
B^\mathrm{NDR}&=&11,\ &B_1^\mathrm{NDR}&=&-\frac{1}{2},\ &B_2^\mathrm{NDR}&=&-\frac{1}{16},\\
B^\mathrm{HV} &=& 7,\ &B_1^\mathrm{HV} &=&0,\            &B_2^\mathrm{HV} &=&-\frac{1}{16}
.\end{array}\end{equation}
This shows the scheme independence of the final result.

  \subsection{Analytical comparison with the literature}
%  \label{sec:checks}
     \label{sec:comparison}
     Next we compare our results with the calculation of the decay rate
$\Gamma_{c\bar cs}$ in \cite{Bagan:1995yf}, where the contributions 
calculated in Section \ref{sec:alphaTree} have been determined already 
in 1995. Using the formulae given in
\cite{Bagan:1995yf} we found, however, that the final result is IR divergent.
Hence, we present here a careful comparison of the individual contributions.
\\
We denote with $D_{xy}$ the diagrams, where the
gluon connects the $x$-quark with the $y$-quark. 
In  \cite{Bagan:1994zd} and \cite{Bagan:1995yf} a different notation
was used, the transcription reads:
\begin{equation}
\begin{array}{llll}
\mbox{Diagram} & \mbox{ \bf VI}   & \leftrightarrow & D_{c \bar{c}} \; ,
\\
\mbox{Diagram} & \mbox{ \bf VIII} & \leftrightarrow & D_{c s      } \; ,
\\
\mbox{Diagram} & \mbox{ \bf X}    & \leftrightarrow & D_{b \bar{c}} \; ,
\\ 
\mbox{Diagram} & \mbox{ \bf XI}   & \leftrightarrow & D_{b s      } \; .
\end{array}
\end{equation}
For the virtual corrections (i.e. the three-particle-cuts in the
notation of \cite{Bagan:1995yf}), the coefficients in front of the 
loop functions could be compared and we found, that the virtual results for 
the diagram $D_{c\bar c}$ 
coincide. 
For the diagrams $D_{cs}$ and $D_{b\bar c}$ 
a factor $-1$ is missing in the results of
\cite{Bagan:1995yf}.
In the diagram $D_{bs}$ the difference is, however, a little more subtle.
Eq.(20) in \cite{Bagan:1995yf}
\begin{equation}
\mathrm{Im}\big[\mathrm{XI}+\mathrm{XI}^\dagger\big]^{(3)}=
\frac{g^2}{192\pi^5b}\int_{4c}^b\frac{\mathrm{d}t}{t}(b-t)^2v
\Big\{(t+2c)b\big
[t(A+4B)+(t+2b)\tilde B\ldots
,\end{equation}
has to be modified to
\begin{equation}
\mathrm{Im}\big[\mathrm{XI}+\mathrm{XI}^\dagger\big]^{(3)}=
\frac{g^2}{192\pi^5b}\int_{4c}^b\frac{\mathrm{d}t}{t}(b-t)^2v
\Big\{(t+2c)b\big[t(A-2B)+2(b-t)\tilde B\ldots
\; .\end{equation}
These two expressions would be equal, if $2 B + \tilde B = 0$ holds,
which is clearly not the case.
For the real corrections (i.e. the four-particle-cuts in the
notation of \cite{Bagan:1995yf}) our results coincide with the 
ones in \cite{Bagan:1995yf} for the diagrams $D_{c\bar c}$ and $D_{cs}$. 
For the two remaining diagrams, no agreement could be found. 
Since in \cite{Bagan:1995yf} these contributions are given in terms of some
phase space functions $\mathcal K_{0\ldots 7}$, while the results obtained 
here are just single expressions, the error could not be traced back easily.

  \subsection{Numerical comparison with the literature}
%  \label{sec:checks}
     \label{sec:comparison-num}
     Using the input parameters from \cite{Bagan:1995yf} for our
newly calculated expressions for $g_{11}$, $g_{12}$ and $g_{22}$, 
we could reproduce all numerical results quoted in Table 2 of 
\cite{Bagan:1995yf}, even though several analytic expressions in 
$g_{12}$ differed, as discussed in detail in Section \ref{sec:comparison}.
Hence we conclude, that the results in
\cite{Bagan:1995yf} contain simply several misprints\footnote{Since the 
authors of \cite{Bagan:1995yf} are not active in HEP anymore, we could
not settle this issue without performing the calculation anew.}.

  \subsection{Numerical results for the decay width}
  \label{sec:numRes}
  For our up-to-date numerical evaluation we used the 
following input parameters from \cite{Lenz:2010gu,pdg2012}:
\begin{equation}
\begin{array}[t]{lcll }
\m_b\leri{\m_b}&= & \leri{4.25\pm0.05}   &\mathrm{GeV} \; ,
\\
\m_s\leri{\m_b}&= & \leri{0.085}   &\mathrm{GeV} \; , 
% In Lenz:2010gu ist dieser Wert gegeben. Die PDG gibt ms(\mu\approx2GeV)=0.95, 
%was zu einem kleineren ms(mb) führt. Numerisch macht das keinen unterschied in der Breite.
\\
\Lambda_{QCD}  &= & \leri{0.213\pm0.008} &\mathrm{GeV} \; ,
\\
|V_{us}|       &= & 0.225\pm0.001\; ,         &
\\
|V_{ub}|       &= & 0.004\pm0.001\; ,         &
\\
|V_{cb}|       &= & 0.041\pm0.001\; ,         &
\\
\delta_{CKM}   &= & 71^\circ\pm25^\circ\; .  &
\end{array}
\end{equation}
The remaining CKM elements were inferred from the unitarity of the 
CKM matrix. 
The decay rate $\Gamma (b \to c \bar{c} s)$  has a strong dependence on the 
charm quark mass, where a lot of progress has been made in its accurate 
determination in recent years. We use the following three 
values \footnote{See also \cite{Alekhin:2012vu} for another recent determination.} for our
numerics:
      \begin{equation}
      \begin{array}[t]{lcll }
        a): \m_c\leri{\m_c}&= & \leri{1.273 \pm 0.006}   &\mathrm{GeV} \;  %cite{McNeile:2010ji}
        ,
        \\
        b): \m_c\leri{\m_c}&= & \leri{1.279 \pm 0.013}   &\mathrm{GeV} \;   %\cite{Chetyrkin:2009fv}
        ,
        \\
       c): \m_c\leri{\m_c}&= & \leri{1.277 \pm 0.026}   &\mathrm{GeV} \;  %\cite{Dehnadi:2011gc}
        .
      \end{array}
      \end{equation}
$a)$ stems from \cite{McNeile:2010ji}, $c)$ from \cite{Chetyrkin:2009fv} and $c)$
from \cite{Dehnadi:2011gc}.
The central values of these three determinations agree excellently, the error estimates
range from $0.5 \%$ \cite{McNeile:2010ji} to $2 \%$ \cite{Dehnadi:2011gc}, which has a visible effect 
in our numerical analysis.
%
%%%%%%%%%%%%%%%%%%%%%%%%%%%%%%%%%%%%%%%%%%%%%%%%%%%%%%%%%%%%%%%%%%%%%%%%%%%%%%%%%%%%%%%%%%%%%%%%%
\begin{table}\centering\begin{tabular}[H]{c|ccccc}
 		&		&	$\delta \mu$	&	$\delta m_c$	&	$\delta m_b$	&	$\delta\Lambda_{QCD}$	\\\hline
$\Gamma^{(0)}_{c\bar cs}$	&	$1.37$	&$\pm0.11$		&$\pm0.07$		&$\pm0.04$		&	$\pm0.02$	 	\\
$\Gamma^{\alpha_s}_{c\bar cs}$	&	$0.34$	&$\pm0.07$		&$\pm0.01$		&$\pm0.00$		&	$\pm0.01$		\\
$\overline\Gamma_{c\bar cs}^m$	&	$0.12$	&$\pm0.03$		&$\pm0.03$		&$\pm0.01$		&	$\pm0.01$		\\
$\Gamma^\mathrm{PO}_{c\bar cs}$	&	$-0.09$	&$\pm0.04$		&$\pm0.00$		&$\pm0.00$		&	$\pm0.00$		\\
$\Gamma^C_{c\bar cs}$		&	$-0.07$	&$\pm0.04$		&$\pm0.00$		&$\pm0.00$		&	$\pm0.00$		\\
$\Gamma^\mathrm{PI}_{c\bar cs}$	&	$-0.05$	&$\pm0.01$		&$\pm0.00$		&$\pm0.00$		&	$\pm0.00$		\\
$\Gamma^{Q_8}_{c\bar cs}$	&	$0.01$	&$\pm0.00$		&$\pm0.00$		&$\pm0.00$		&	$\pm0.00$		\\\hline
$\Gamma_{c\bar cs}$		&	$1.63$	&$\pm0.15$		&$\pm0.10$		&$\pm0.04$		&	$\pm0.03$
\end{tabular}\caption{Single contributions to the $c\bar cs$-decay width in units of $\leri{\Gamma_0|V_{cb}|^2}$ and the corresponding errors.}\label{tab:contributions}\end{table}
With these inputs, 
the following value for the total width is obtained:
\begin{eqnarray}
\Gamma_{c\bar cs}^{a)}&=&
\Gamma_0 |V_{cb}|^2 
\leri{1.64\pm 0.15_\mu \pm 0.04_{m_b}\pm 0.02_{m_c} \pm0.03_{\Lambda_{QCD}}}
 %\cite{McNeile:2010ji}
,
%\\
% &= &
%\Gamma_0 V_{cb}^2 \leri{1.71\pm0.33}%MS-schema%Davies
\\
\Gamma_{c\bar cs}^{b)} &=&
\Gamma_0 |V_{cb}|^2
\leri{1.62\pm0.15_\mu  \pm0.04_{m_b}\pm0.05_{m_c} \pm0.03_{\Lambda_{QCD}}}%MS-schema%Steinhauser
%\cite{Chetyrkin:2009fv}
,
\\
\Gamma_{c\bar cs}^{c)}&=&
\Gamma_0 |V_{cb}|^2
\leri{1.63\pm0.15_\mu \pm0.04_{m_b}\pm0.10_{m_c}   \pm0.03_{\Lambda_{QCD}}}
%\cite{Dehnadi:2011gc}
.
\label{Gamma_res}
\end{eqnarray}
All three determinations agree perfectly, the only sizable difference is in the
theoretical error due to the charm quark mass, which ranges from $1\%$
to $6\%$. Since the factor $\Gamma_0 |V_{cb}|^2$ cancels in most branching ratios 
exactly, we did not include this term in the error analysis.
The individual contributions to the inclusive decay width are given in 
Table \ref{tab:contributions}, where also all parametric uncertainties are listed. These 
results will be discussed below in detail.
\\
As the main numerical result of our paper we present new values for the branching ratio
of the inclusive $b \to c \bar{c} s$ transition.
\begin{eqnarray}
Br(b \to c \bar{c} s) &=& \frac{\Gamma_{c\bar cs}}{\Gamma_{tot}} \, .
\end{eqnarray}
For the total decay rate $\Gamma_{tot}$
we took all theoretical expressions that were available in the literature,
$b \to c l^- \bar{\nu}$ from \cite{Nir:1989rm},
$b \to c \bar{u} d    $ from \cite{Bagan:1994zd},
$b \to c \bar{c} s    $ from this work,
$b \to $ no charm       from \cite{Lenz:1997aa}
and 
$b \to s g $            from \cite{Greub:2000sy,Greub:2000an}.
We finally get
\begin{eqnarray}
Br(b \to c \bar{c} s)^{a)} &=&
0.234  \pm 0.002_\mu \pm 0.003_{m_b} \pm 0.001_{m_c} \pm 0.001_{\Lambda_{QCD}}
 %\cite{McNeile:2010ji}
,
\nonumber
\\ &&
\\
Br(b \to c\bar{c} s)^{b)} &=&
0.232  \pm 0.002_\mu \pm 0.003_{m_b} \pm 0.003_{m_c} \pm 0.001_{\Lambda_{QCD}}
%MS-schema%Steinhauser
%\cite{Chetyrkin:2009fv}
,
\nonumber
\\ &&
\\
Br(b \to c\bar{c} s)^{c)} &=&
0.232  \pm 0.002_\mu \pm 0.003_{m_b} \pm 0.006_{m_c} \pm 0.001_{\Lambda_{QCD}}
%\cite{Dehnadi:2011gc}
.
\nonumber
\\ &&
\end{eqnarray}
The errors in the branching ratios are considerably smaller than in the decay rate, 
obviously several uncertainties cancel to some extend in the branching ratios.
Our final results show several interesting features
(to be conservative we will use in the following the value of \cite{Dehnadi:2011gc}
for the charm quark mass):
\begin{itemize}
\item All NLO-QCD corrections enhance the LO-QCD result for the decay rate by $+19 \%$. 
      The contribution of the 1-loop gluon correction to the insertion of the operators
      $Q_1$ and $Q_2$ is even $+25\%$ of the LO-QCD result.
      As already pointed out in \cite{Voloshin:1994sn,Bagan:1995yf}, effects of a finite
      charm quark mass were crucial. This can be seen if one compares our results with 
      the one for a vanishing charm quark mass:
      \begin{equation}
      \Gamma_{c\bar cs}^{\leri{\overline m_c=0}} =
      \Gamma_0 |V_{cb}|^2 
       \leri{3.17^{(0)} 
            \! -0.07^{\alpha_s} 
            \! +\overline{0.37}^m
            \! \! -0.14^\mathrm{PO}
            \! \! -0.16^C
            \! \! -0.08^\mathrm{PI}
            \! +0.03^{Q_8}} \; .
      \end{equation}
      Because of the missing phase space suppression the LO contribution is now more than
      a factor 2 larger than in Table \ref{tab:contributions}. The biggest dependence is found,
      however, in $\Gamma^{\alpha_s}_{c\bar cs}$, which changes from $-0.07 \; \Gamma_0 |V_{cb}|^2$ 
      in the massless case to $+0.34  \; \Gamma_0 |V_{cb}|^2$ in the massive case.
\item All penguin effects give a contribution of about $- 9\% $ of the LO-QCD decay rate.
      The newly calculated penguin insertions are of a similar size as some of the theoretical
      uncertainties, so their inclusion is reasonable.
      Gluon corrections to the insertion of the penguin operators $Q_3,...,Q_6$ in 
      tree level diagrams of the effective theory are still missing. 
      Naively one expects these corrections to be of the order of
      \begin{displaymath}
      \frac{\Gamma_{c \bar{c} s}^{\alpha_s} }{\Gamma_{c \bar{c} s}^{(0)}}
           \Gamma_{c \bar{c} s}^{PO}
      \approx \frac{0.34}{1.37} \; 0.09 \approx 0.02 \; .
      \end{displaymath}
      This is much smaller than the parametric uncertainties of our results. 
      Hence, we consider the effort of the corresponding NLO-QCD calculation not to be justified.
\item The dominant theoretical uncertainty in our above analysis stems from the 
      renormalisation scale dependence. 
      It is of the order of $\pm 9 \%$ of the decay rate and cancels in
      the branching ratio to a remaining uncertainty of the order 
      of $\pm 1 \%$.
      To reduce this uncertainty further a NNLO-QCD calculation would be mandatory.
      The dependence on the values of the charm quark mass and the bottom quark mass
      is already subleading, as well as the dependence on the strong coupling.
      \\
      The dominant dependence of the decay rate on the CKM elements is given by 
      the overall factor $|V_{cb}|^2$, which 
      results in an uncertainty of about $5 \%$. 
      Since we are interested in the end in experimentally measurable branching ratios
      we did not include the prefactor $\Gamma_0 |V_{cb}|^2$, which cancels exactly in the ratios,
      in the error analysis of the decay rate.
      The remaining CKM dependence is  negligible.
      \\
      We also investigated the effect of a non-vanishing strange quark mass and got
      for $\overline m_s\leri{\overline m_b}=0.085 \, \mathrm{GeV}$
      \begin{equation}
      \Gamma_{c\bar cs}=
      \Gamma_0 |V_{cb}|^2 
      \leri{1.62\pm0.15_\mu \pm0.04_{m_b}\pm0.10_{m_c}   \pm0.03_{\Lambda_{QCD}}} \; ,
      \end{equation}
      which is almost  equivalent to Eq.(\ref{Gamma_res}). So this effect can be safely neglected.
\item In \cite{Beneke:2002rj} it was shown that using $\overline{m}_c(\overline{m}_b)$ instead of
      $\overline{m}_c(\overline{m}_c)$ sums up large logarithms of the form $x_c^2 \ln x_c^2$ to all orders.
      We use this prescription also in this work, which also solves a second issue:
      one might consider the natural scale of the decay $b \to c \bar{c} s$ to be
      $\sqrt{m_b^2 - 4 m_c^2}$. Using our numerical input we get for 
      the renormalisation scale 
      $\mu = \sqrt{1-4 (\overline{m}_c(\overline{m}_b)m_c/\overline{m}_b(\overline{m}_b))^2} 
      \; m_b \approx 0.9 \; m_b$,
      which is very close to our choice $\mu = m_b$. Thus we see no reason for choosing 
      different renormalisation scales in different $b$-decay channels.
      Choosing different scales would enhance the theoretical uncertainties
      in the branching ratios sizable.
\item Since we are claiming a high precision of our final result, hypothetical drawbacks
      of our theoretical tools have to be investigated in detail.
      For the calculation of inclusive decay rates within the HQE, one has to take in to account 
      {\bf all} possible cuts through the corresponding forward-scattering 
      diagrams. In the case of penguin insertions and contributions from $Q_8$, some cuts, however,
      do not belong to $b \to c \bar{c} s$, but to the decay $b \to s g$, see Fig.(\ref{pic:analog}).
      \begin{figure}[h]
      \centering
      \fbox{
      \includegraphics[width=0.7\textwidth]{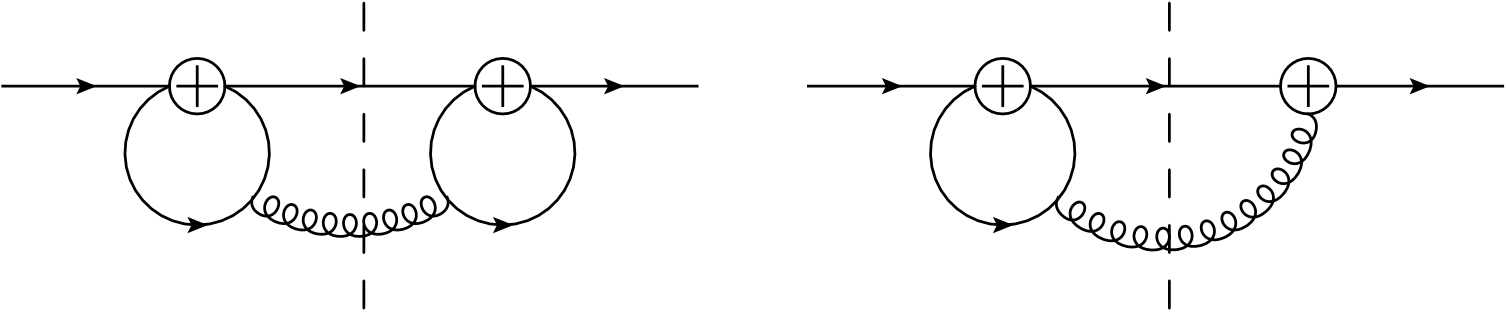}
            }
      \caption{Cuts through forward scattering diagrams related to penguin insertions 
         and $Q_8$-contributions, which belong not to $b \to c \bar{c} s$ but to the $b \to sg$ decay.
           }
      \label{pic:analog}
      \end{figure}
      Such a feature could in principle spoil the application of the HQE to the decay $b \to c \bar{c} s$. 
      All these cuts however 
      involve a quark-loop from which a real gluon is emitted. The corresponding matrix element can 
      be expressed as:
      \begin{equation}\begin{aligned}\label{eq:cuts}
      \mathcal M \propto \leri{m_l^2 B_0+ \leri{2-D}B_{00}}\bar s\slashed\epsilon^*\leri{1-\gamma_5}b\\
      +\leri{B_1+B_{11}}\leri{2-D}\bar s\slashed p_g\slashed\epsilon^*\slashed p_g\leri{1-\gamma_5}b
      ,\end{aligned}\end{equation}
      where the $B_{ij}$ are the two-point loop-integrals and $p_g^\mu$ and $\epsilon^\mu$ 
      are the gluon's momentum- and polarisation vector. $m_l$ is the mass of the particle 
      running in the loop. The second line of (\ref{eq:cuts}) vanishes due to the on-shell 
      condition of the gluon: $p_g^\mu p_{g\mu}=p_g^\mu \epsilon^*_\mu=0$.\\
      For a vanishing momentum-square of the gluon, the remaining loop-functions, 
      $B_0$ and $B_{00}$, are related by:
      \begin{equation}
      B_{00}=\frac{m_l^2}{2}\frac{1}{1-\varepsilon}B_0
      ,\end{equation}
      in $D=4-2\varepsilon$ dimensions. With this relation, also the first line 
      of (\ref{eq:cuts}) vanishes and these dangerous cuts do not contribute.
\item Finally we compare the numerical values for the
      branching ratio of $b \to c \bar{c} s$ for different 
      schemes of the $b$-quark mass.
      We use the pole scheme, 
      the $\overline{MS}$-scheme \cite{Bardeen:1978yd},
      the kinetic                   \cite{Bigi:1996si},
      the potential-subtracted (PS) \cite{PSMass} 
      and the $\Upsilon(1s)$-scheme \cite{1SMass}.
      Our final results in these schemes reads:
      \begin{eqnarray}
      \mathcal B_{c\bar cs}^\mathrm{Pole} & = &0.166  \pm 0.011 \; ,
      \\
      \mathcal B_{c\bar cs}^{\overline{\mathrm{MS}}}& = &0.232  \pm 0.007\; ,
      \\
      \mathcal B_{c\bar cs}^\Upsilon & = &0.243  \pm 0.013\; ,
      \\
      \mathcal B_{c\bar cs}^\mathrm{PS} & = &0.241  \pm 0.013\; ,
      \\
      \mathcal B_{c\bar cs}^\mathrm{KIN} & = &0.245  \pm 0.013\; .
      \end{eqnarray}
      Except for the pole mass scheme all other quark mass schemes agree very
      nicely, the largest shift is found in the kinetic scheme, which is
      about $0.013$, i.e. $ 6\%$ larger than the result in the $\mathrm{MS}$ scheme.
      Since the pole scheme has theoretical disadvantages, as discussed in
      Section \ref{sec:MSbar}, we will not use the result in this scheme.
      The numerical difference in the remaining four schemes will be used as
      an estimate for an additional systematic uncertainty, which we estimate to be
      $\pm 0.013$.
\end{itemize}
Thus we get as our final result for the branching ratio      
\begin{eqnarray}
      \mathcal B_{c\bar cs} & = & 0.232  \pm 0.007 \pm 0.013
                            \approx (23 \pm 2) \% \; .
      \end{eqnarray}

\section{Conclusion}
\label{sec:conclusion}
In this work, the width for the inclusive decay 
channel $b\rightarrow c\bar cs$ 
was computed up to the order $\ord{\alpha_s}$ within the framework of the 
HQE.
This theoretical framework passed recently a non-trivial
experimental test: the measured value of the decay rate 
difference $\Delta \Gamma_s$ in the neutral $B_s$ meson system
agrees nicely with the corresponding standard model
prediction \cite{Lenz:2012mb}. 
Another very non-trivial test is related to the huge
lifetime difference in the $D$-meson systems. Here first results, 
although suffering from huge hadronic uncertainties, look very
promising \cite{Lenz:2013aua}.
Thus an updated prediction of $\Gamma (b\rightarrow c\bar cs)$
within the framework of the HQE is clearly overdue. 
\\
For that purpose we had to recalculate radiative corrections to 
insertions of the
dominant operators $Q_1$ and $Q_2$ in tree level diagrams of the 
effective theory. This task was already performed in \cite{Bagan:1995yf},
but the formul\ae\  in  \cite{Bagan:1995yf} contain several misprints.
We give the corrected expressions in the appendix of this paper.
However, numerical results of \cite{Bagan:1995yf} could be exactly 
reproduced.
In addition to that we also determined for the first time
NLO-QCD contributions from insertions of the operators 
$Q_1$ and $Q_2$ in penguin diagrams of the 
effective theory, as well as
contributions from the chromomagnetic operator $Q_8$. 
\\
Combining our new calculation with the impressive improvements
in the accurate determination of standard model parameters
like quark masses and CKM parameters we
obtain as the main numerical result a very precise 
value of  the branching ratio of the decay 
$b \to c \bar{c} s$
\begin{equation}
Br (b \to c \bar{c} s) = (23 \pm 2) \% \; .
\end{equation}
Further support for this small error is given by new insights in 
different quark mass schemes.
\\
With this new result a re-analysis of the semileptonic 
branching-ratio $\mathcal B_{Sl}$ and the charm-multiplicity $n_c$ 
can be performed, see \cite{planned}. Such an analysis will lead to 
important and complementary insights on the possible size of new
physics effects in $B$ decays.

\section*{Acknowledgments}
We would like to thank M. Gorbahn, A. Kagan, G. Kirilin, U. Nierste 
and J. Rohrwild for enlightening discussions; C. Davies, A. Hoang and
M. Steinhauser for helpful comments and also E. Bagan and B. Fiol for 
trying to dig out their old computer programs. 
F.K. and T.R. would like to thank the IPPP Durham and KIT Karlsruhe 
for hospitality. In addition, F.K. and T.R. would like to thank 
M. Beneke and A. Ibarra for support during their thesis.

\appendix
\section{An introduction to \cite{hokimpham1984}}
\label{sec:pedestrian}
The corrections to the tree level-diagrams, that involve gluons coupling to only one fermion-line were given in the paper \cite{hokimpham1984} in terms of a one dimensional integral. For example the term $g_{22}$ can be expressed as:
\begin{equation}
g_{22}=\frac{4}{\Omega_0}\leri{\Gamma_l+\Gamma_u}
.\end{equation}
Here, as in this whole section, the notation of \cite{hokimpham1984} is used. The two contributions $\Gamma_l$ and $\Gamma_u$ depend on the mass ratios $\rho_i=\leri{m_i/M}^2$. $\Gamma_u$ hereby contains all effects due to gluon-exchanges on the upper vertex, the vertex with the decaying particle incoming and one particle outgoing. In the $b\to c\bar cs$-case, the upper vertex is the $b$-$c$-$W^-$-vertex. $\Gamma_l$ contains the corrections from gluon exchanges on the lower vertex,  which in this case is the $\bar c$-$s$-$W^-$-vertex. As already mentioned, diagrams that contain gluon-exchanges between upper and lower vertex vanish due to the colour-structure. The factor $\Omega_0$ is given by equation (3.9) in \cite{hokimpham1984}.\\
As usual for such radiative corrections, the $\Gamma_{u,l}$ each consist of a part describing virtual- and real-gluon-effects respectively:
\begin{eqnarray}
\Gamma_u&=&\Gamma_l^v+\Gamma_l^b,\\
\Gamma_l&=&\Gamma_u^v+\Gamma_u^b
,\end{eqnarray}
which are defined in equations (3.36) and (4.34) as one dimensional integrals over the functions $R_{l,u}^{b,v}\leri{\xi}$ respectively. These are given in equations (3.11), (3.35), (4.7) and (4.33).

\section{Expressions for the radiative corrections}
\label{sec:radAn}
The expressions for the $\ord{\alpha_s}$ corrections to $b\rightarrow c\bar cs$ are split in a virtual part, e.g. $g_{c\bar c}^\mathrm{v}$, stemming from loop-corrections, and a real part $g_{c\bar c}^\mathrm{r}$ from gluon radiation. The sums of both contributions, $g_{x}=g_{x}^\mathrm{v}+g_{x}^\mathrm{r}$, are infrared-finite for every diagram, as expected.
\subsection{Expressions for the virtual contributions}
The virtual correction $g_{c\bar c}^\mathrm v$ reads with the infrared-regulator $\xi=m_g/m_b$:
\begin{equation}\begin{array}[t]{c}
g_{c\bar c}^\mathrm{v}=\int_{4x_c^2}^1dz_{12}\frac{-4}{z_{12}^{3/2}}\leri{1-z_{12}}^2\sqrt{z_{12}-4x_c^2}\Bigg[z_{12}\leri{1+2z_{12}}\big[-2\leri{\C_\mathtt{00}+B_1}\\
+\C_\mathtt0z_{12}+\C_\mathtt1 z_{12}+\C_\mathtt{12}z_{12}+\C_\mathtt2 z_{12}\big]+2x_c^4\big[-2\C_\mathtt2 +2\C_\mathtt0\leri{z_{12}-1}\\
+2\C_\mathtt1\leri{z_{12}-1}+3\C_\mathtt{11}z_{12}+9\C_\mathtt{12}z_{12}+2\C_\mathtt2z_{12}\big]\\
+x_c^2\Big[4\leri{\C_\mathtt{00}+B_1}\leri{z_{12}-1}-\big[3\C_\mathtt1+4\C_\mathtt{12}+3\C_\mathtt2\big]z_{12}\\
-\big[6\C_\mathtt0+9\C_\mathtt1+3\C_\mathtt{11}+11\C_\mathtt{12}+9\C_\mathtt2\big]z_{12}^2\Big]\Bigg]
.\end{array}\label{eq:virtuellVI}\end{equation}
The functions $\C_\mathtt{ij}$ are the three-point-functions as used in the \textit{LoopTools}-package \cite{hahnperez1999}, their arguments read in the notation adapted in the \textit{LoopTools}-manual: $p_1^2=(p_1+p_2)^2=m_2^2=m_3^2=x_c^2$, $p_2^2=z_{12}$ and $m_1^2=\xi^2$. The arguments are chosen in a way that renders the loop-functions dimensionless. Therefore, also the renomalization-point is set to $\mu^2/m_b^2$.

The function $g_{cs}^\mathrm v$ is given by:
\begin{equation}\begin{array}[t]{c}
g_{cs}^\mathrm{v}=\int_{x_c^2}^{\leri{1-x_c}^2}dz_{12}\frac{-12}{z_{12}}\leri{z_{12}-x_c^2}^2\leri{1+x_c^2-z_{12}}\\
\sqrt{x_c^4+\leri{1-z_{12}}^2-2x_c^2\leri{1+z_{12}}}\Big[16\leri{\C_\mathtt{00}+B_2}+4\C_\mathtt1x_c^2+4\C_\mathtt{11}x_c^2\\
+4\C_\mathtt{12}x_c^2+2\C_\mathtt2x_c^2+2\C_\mathtt0\leri{x_c^2-z_{12}}-2\C_\mathtt1z_{12}-4\C_\mathtt{12}z_{12}-2\C_\mathtt2z_{12}\Big]
.\end{array}\label{eq:virtuellVIII}\end{equation}
Here the arguments of the \textit{LoopTools}-functions are given by: $p_1^2=m_2^2=x_c^2$, $(p_1+p_2)^2=m_3^2=0$, $p_2^2=z_{12}$ and $m_1^2=\xi^2$.

The third virtual correction is given by:
\begin{equation}\begin{array}[t]{c}
g_{b\bar c}^\mathrm{v}=\int_{x_c^2}^{\leri{1-x_c}^2}\frac{-12}{z_{12}}\leri{z_{12}-x_c^2}^2\sqrt{x_c^4+\leri{1-z_{12}}^2-2x_c^2\leri{1+z_{12}}}\\
\Big[4\C_\mathtt{1}+4\C_\mathtt{11}+4\C_\mathtt{12}+2\C_\mathtt{2}+2\C_\mathtt1x_c^2+4\C_\mathtt{11}x_c^2+12\C_\mathtt{12}x_c^2+2\C_\mathtt2x_c^2+2\C_\mathtt1x_c^4\\
+8\C_\mathtt{12}x_c^4+4\C_\mathtt2x_c^4+16\leri{\C_\mathtt{00}+B_2}\leri{1+x_c^2-z_{12}}+2\C_\mathtt0\leri{1+x_c^2-z_{12}}^2\\
-6\C_\mathtt1z_{12}-4\C_\mathtt{11}z_{12}-8\C_\mathtt{12}z_{12}-4\C_\mathtt2z_{12}-4\C_\mathtt1x_c^2z_{12}-12\C_\mathtt{12}x_c^2z_{12}\\
-6\C_\mathtt2x_c^2z_{12}+2\C_\mathtt1z_{12}^2+4\C_\mathtt{12}z_{12}^2+2\C_\mathtt2z_{12}^2\Big]
.\end{array}\label{eq:virtuellX}\end{equation}
The three-point-functions' arguments are: $p_1^2=m_2^2=1$, $(p_1+p_2)^2=m_3^2=x_c^2$, $p_2^2=z_{12}$ and $m_1^2=\xi^2$.

The last virtual correction reads:
\begin{equation}\begin{array}[t]{c}
g_{bs}^\mathrm{v}=\int_{4x_c^2}^{1}dz_{12}\frac{-4}{z_{12}^{3/2}}\leri{1-z_{12}}^2\sqrt{z_{12}-4x_c^2}\Big[-2x_c^2\big[2\leri{\C_\mathtt{00}+B_1}+\C_\mathtt1+\C_\mathtt2\\
+\C_\mathtt0\leri{1-z_{12}}^2-2\leri{\C_\mathtt{00}+B_1}z_{12}-2\C_\mathtt1z_{12}-3\C_\mathtt{11}z_{12}-\C_\mathtt{12}z_{12}-2\C_\mathtt2z_{12}\\
+\C_\mathtt1z_{12}^2+\C_\mathtt{12}z_{12}^2+\C_\mathtt2z_{12}^2\big]+z_{12}\big[-2\leri{\C_\mathtt{00}+B_1}-\C_\mathtt1-\C_\mathtt2-4\leri{\C_\mathtt{00}+B_1}z_{12}\\
-4\C_\mathtt1z_{12}-3\C_\mathtt{11}z_{12}-2\C_\mathtt{12}z_{12}-\C_\mathtt2z_{12}+2\C_\mathtt1z_{12}^2+2\C_\mathtt{12}z_{12}^2+2\C_\mathtt{2}z_{12}^2\\
+\C_\mathtt0\leri{2z_{12}^2-z_{12}-1}\big]\Big]
.\end{array}\label{eq:virtuellXI}\end{equation}
The arguments of the loop-functions are: $p_1^2=m_2^2=1$, $(p_1+p_2)^2=m_3^2=0$, $p_2^2=z_{12}$ and $m_1^2=\xi^2$.\\
Like in section \ref{sec:scheme}, the two coefficients $B_{1,2}$ encode the scheme-dependence of the results. They always appear in the combination $\C_\mathtt{00}+B_{1,2}$, since $\C_\mathtt{00}$ is the only UV-divergent integral. In the NDR- and HV-scheme, these coefficients are given by:
\begin{equation}
B_1^\mathrm{NDR}=-\frac{1}{2},\ B_2^\mathrm{NDR}=-\frac{1}{16},\ B_1^\mathrm{HV}=0,\ B_2^\mathrm{HV}=-\frac{1}{16}
.\end{equation}
\subsection{Expressions for the real parts}
The corresponding real corrections read, again with $\xi=m_g/m_b$ as an infrared regulator:
\begin{equation}\begin{array}[t]{c}
g_{c\bar c}^{\mathrm{r}}=\int_{\leri{2x_c+\xi}^2}^{1}dz_{123}\int_{\leri{x_c + \xi}^2}^{\leri{\sqrt{z_{123}}-x_c}^2}dz_{12} (-6) \leri{z_{123}-1}^2\Bigg[-\leri{z_{12}-x_c^2}\\
\Big[x_c^4+\leri{z_{12}-z_{123}}^2-2x_c^2 \leri{z_{12}+z_{123}}\Big]^{5/2}\Big[-3x_c^4\leri{2+z_{123}}-z_{12}^2\leri{2+z_{123}}\\
+z_{12}z_{123}\leri{2+z_{123}}+2z_{123}^2 \leri{1+2z_{123}}+x_c^2\big(-3z_{123}^2+4z_{12}\leri{2+z_{123}}\big)\Big]\\
+2z_{12}\big(x_c^4+\leri{z_{12}-z_{123}}^2-2x_c^2\leri{z_{12}+z_{123}}\big)^2\Big[x_c^6\leri{2+z_{123}}\\
-\leri{z_{12}-z_{123}}z_{123}^2\leri{1+2z_{123}}-x_c^4\big[-3\leri{z_{123}-1}z_{123}+2z_{12}\leri{2+z_{123}}\big]\\
+x_c^2\big(z_{12}\leri{z_{123}-1}z_{123}+z_{123}^2-4z_{123}^3+z_{12}^2\leri{2+z_{123}}\big)\Big]\\
\bigg[\ln\bigg(-x_c^4+z_{12}\leri{z_{12}-z_{123}}-\leri{z_{12}+z_{123}}\xi^2+x_c^2\leri{z_{123}+\xi^2}-\\
\sqrt{x_c^4+\leri{z_{12}-z_{123}}^2-2x_c^2\leri{z_{12}+z_{123}}}\sqrt{x_c^4+\leri{z_{12}-\xi^2}^2-2x_c^2\leri{z_{12}+\xi^2}}\bigg)\\
-\ln\bigg(-x_c^4+z_{12}\leri{z_{12}-z_{123}}-\leri{z_{12}+z_{123}}\xi^2+x_c^2\leri{z_{123}+\xi^2}+\\
\sqrt{x_c^4+\leri{z_{12}-z_{123}}^2-2x_c^2\leri{z_{12}+z_{123}}}\sqrt{x_c^4+\leri{z_{12}-\xi^2}^2-2x_c^2\leri{z_{12}+\xi^2}}\bigg)\bigg]\Bigg]\\
\bigg/\bigg[3\leri{x_c^2-z_{12}}z_{12}z_{123}^3\big(x_c^4+\leri{z_{12}-z_{123}}^2-2x_c^2\leri{z_{12}+z_{123}}\big)^2\bigg]
,\label{eq:reellVI}\end{array}\end{equation}
\begin{equation}\begin{array}[t]{c}
g_{cs}^{\mathrm{r}}=\int_{\leri{x_c+\xi}^2}^{\leri{1-x_c}^2}dz_{123}\int_{\leri{x_c + \xi}^2}^{z_{123}}dz_{12}24\leri{1+x_c^2-z_{123}}\leri{z_{123}-z_{12}}\\
\sqrt{x_c^4+\leri{z_{123}-1}^2-2x_c^2\leri{1+z_{123}}}\Bigg[\leri{x_c^2-z_{12}}\leri{x_c^2-z_{12}+z_{123}}\\
+z_{12}\leri{z_{123}-x_c^2}\bigg[\ln\bigg(1+\frac{\leri{z_{123}-z_{12}}\sqrt{x_c^4+\leri{z_{12}-\xi^2}^2-2x_c^2\leri{z_{12}+\xi^2}}}{\leri{x_c^2-z_{12}}\leri{z_{12}-z_{123}}+\leri{z_{12}+z_{123}}\xi^2}\bigg)\\
-\ln\bigg(1-\frac{\leri{z_{123}-z_{12}}\sqrt{x_c^4+\leri{z_{12}-\xi^2}^2-2x_c^2\leri{z_{12}+\xi^2}}}{\leri{x_c^2-z_{12}}\leri{z_{12}-z_{123}}+\leri{z_{12}+z_{123}}\xi^2}\bigg)\bigg]\Bigg]\\
\bigg/\Big[\leri{x_c^2-z_{12}}z_{12}z_{123}\Big]
,\end{array}\label{eq:reellVIII}\end{equation}
\begin{equation}\begin{array}[t]{c}
g_{b\bar c}^{\mathrm{r}}=\int_{\leri{x_c+\xi}^2}^{\leri{1-x_c}^2}dz_{123}\int_{x_c^2}^{\leri{\sqrt{z_{123}}-\xi}^2}dz_{12}24\leri{x_c^2-z_{12}}^2\\
\Bigg[\big(1+x_c^4+z_{12}\leri{z_{123}-2}-x_c^2\leri{z_{12}+z_{123}-2}\big)\ln\bigg(z_{123}^2+\leri{x_c^2-1}\\
\leri{z_{12}-\xi^2}-z_{123}\leri{x_c^2+z_{12}+\xi^2-1}-\sqrt{x_c^4+\leri{z_{123}-1}^2-2x_c^2\leri{1+z_{123}}}\\
\sqrt{\leri{z_{12}-z_{123}}^2-2\leri{z_{12}+z_{123}}\xi^2+\xi^4}\bigg)+\frac{1}{z_{123}}\bigg[-\leri{z_{12}-z_{123}}^2\\
\sqrt{x_c^4+\leri{z_{123}-1}^2-2x_c^2\leri{1+z_{123}}}+z_{123}\bigg[\big(-1-x_c^4+2x_c^2\leri{z_{12}-1}+z_{12}\\
+z_{123}-z_{12}z_{123}\big)\ln\bigg(z_{123}+z_{12}\leri{z_{123}+x_c^2-1}+\xi^2+z_{123}\leri{\xi^2-z_{123}}\\
-x_c^2\leri{z_{123}+\xi^2}-\sqrt{x_c^4+\leri{z_{123}-1}^2-2x_c^2\leri{1+z_{123}}}\\
\sqrt{\leri{z_{12}-z_{123}}^2-2\leri{z_{12}+z_{123}}\xi^2+\xi^4}\bigg)-\big(1+x_c^4+z_{12}\leri{z_{123}-2}\\
-x_c^2\leri{z_{123}+z_{12}-2}\big)\ln\bigg(z_{12}\leri{x_c^2-1-z_{123}}+z_{123}\leri{1+z_{123}-x_c^2}+\xi^2\\
-\leri{x_c^2+z_{123}}\xi^2+\sqrt{x_c^4+\leri{z_{123}-1}^2-2x_c^2\leri{1+z_{123}}}\\
\sqrt{\leri{z_{12}-z_{123}}^2-2\leri{z_{12}+z_{123}}\xi^2+\xi^4}\bigg)+\big(x_c^4-2x_c^2\leri{z_{12}-1}+\leri{z_{12}-1}\\
\leri{z_{123}-1}\big)\ln\bigg(z_{123}+z_{12}\leri{x_c^2+z_{123}-1}+\xi^2+z_{123}\leri{\xi^2-z_{123}}\\
-x_c^2\leri{z_{123}+\xi^2}+\sqrt{x_c^4+\leri{z_{123}-1}^2-2x_c^2\leri{1+z_{123}}}\\
\cdot\sqrt{\leri{z_{12}-z_{123}}^2-2\leri{z_{12}+z_{123}}\xi^2+\xi^4}\bigg)\bigg]\bigg]\Bigg]\bigg/\Big[z_{12}\leri{z_{12}-z_{123}}\Big]
,\end{array}\label{eq:reellX}\end{equation}
\begin{equation}\begin{array}[t]{c}
g_{bs}^{\mathrm{r}}=\int_{4x_c^2}^{\leri{1-\xi}^2}dz_{123}\int_{4x_c^2}^{z_{123}}dz_{12}6\sqrt{z_{12}-4x_c^2}\leri{z_{12}-z_{123}}\\
\Bigg[\Big[z_{12}\big(z_{12}-4z_{12}^2-z_{12}z_{123}+\leri{z_{123}-3}z_{123}\big)+2x_c^2\big(z_{12}+2z_{12}^2-z_{12}z_{123}\\
+\leri{z_{123}-3}z_{123}\big)\Big]\leri{1-z_{123}}-2\leri{z_{12}-1}\big(-2x_c^2\leri{z_{12}-1}+z_{12}\leri{1+2z_{12}}\big)\\
z_{123}\bigg[\ln\bigg(z_{12}\Big(-1+z_{123}+\xi^2-\sqrt{\leri{z_{123}-1}^2-2\leri{1+z_{123}}\xi^2+\xi^4}\Big)\\
+z_{123}\Big(1-z_{123}+\xi^2+\sqrt{\leri{z_{123}-1}^2-2\leri{1+z_{123}}\xi^2+\xi^4}\Big)\bigg)\\
-\ln\bigg(z_{123}-z_{123}\Big(z_{123}-\xi^2+\sqrt{\leri{z_{123}-1}^2-2\leri{1+z_{123}}\xi^2+\xi^4}\Big)\\
+z_{12}\Big(z_{123}-1+\xi^2+\sqrt{\leri{z_{123}-1}^2-2\leri{1+z_{123}}\xi^2+\xi^4}\Big)\bigg)\bigg]\Bigg]\\\bigg/\Big[3z_{12}^{3/2}\leri{z_{123}-1}z_{123}\Big]
.\end{array}\label{eq:reellXI}\end{equation}

\section{Phase space factors for the penguin-insertion}
\label{sec:pengAn}
The phase-space-integrals from the penguin insertion of $Q_2$ are, with $y_c=\sqrt{1-4x_c^2}$, given by:
\begin{equation}\begin{array}[t]{c}
g_{\mathrm{PI}}=-\frac{1}{9}\Bigg\{3 i \pi  x_c^2 \big[-15+8 x_c^2 \leri{9-20 x_c^2+8 x_c^4}+24 x_c^2 \leri{4 x_c^2-1} \Ln{4 x_c^2}\big]\\
+\frac{1}{4 y_c}\bigg[19-30 x_c^2-330 x_c^4+476 x_c^6+432 x_c^8+144 x_c^4 y_c \leri{4+3 x_c^4} \\
\Ln{\frac{2 x_c}{1+y_c}}+96 x_c^2 y_c \leri{1+20 x_c^4}\Ln{\frac{1+y_c}{2 x_c}}\bigg]\\
+3 \leri{1-4 x_c^2+36 x_c^4-48 x_c^6} \Ln{\frac{-1+2 x_c^2+y_c}{2 x_c^2}}\\
-4 \bigg[12 x_c^2 \Ln{\frac{2 x_c}{1+y_c}} \Big[1-2 x_c^2 \leri{2+3 \leri{x_c^2+x_c^4}}+6 x_c^2 \leri{x_c^4-1} \Ln{\frac{x_cm_b}{\mu}}\Big]\\
-\frac{1}{2} y_c \Big[1-44 x_c^2+214 x_c^4-36 x_c^6+\leri{-3+6 x_c^2 \leri{7+x_c^2+6 x_c^4}}\Ln{\frac{x_cm_b}{\mu}}\Big]\bigg]\\
+24 x_c^4 \leri{4 x_c^2-1} \bigg[\pi  \leri{\pi +6 i \Ln{x_c}}+6 i \Ln{\frac{i\leri{1+y_c}}{2x_c}} \\
\Big[
-i \Ln{\frac{i\leri{1+y_c}}{2x_c}}
-2 i \Ln{\frac{1- y_c}{2 x_c^2}}
\Big]
-6 \Li{-\frac{1-2 x_c^2- y_c}{2 x_c^2}}
\bigg]
\Bigg\}
,\label{eq:gc224}\end{array}\end{equation}
\begin{equation}\begin{array}[t]{c}
g_{\mathrm{PI}+}=-\frac{2 x_c^2 }{3} \Bigg\{5 y_c+x_c^2 \Big[-47 y_c+2 \pi^2 \leri{8 x_c^2-1}+4 i \pi  \leri{40 x_c^4-9}+\\
x_c^2 \big[-78 y_c+480 \Ln{2}^2\big]-6 \Ln{2} \leri{3+10\Ln{2}}\Big]+24 x_c^2 \leri{-1+8 x_c^2}\cdot\\
 \Ln{x_c}\Ln{16 x_c}+24 x_c^2 \leri{8 x_c^2-1}\Ln{2 x_c} \big[-\Ln{1-y_c}+\Ln{1+y_c}\big]\\
+2 \Ln{\frac{-1+2 x_c^2+y_c}{2 x_c^2}}-4 y_c \Ln{\frac{x_c m_b}{\mu}}+2 x_c^2 \bigg[-12 i \pi  \leri{8 x_c^2-1}\Ln{2 x_c}\\
+6 \Ln{1+y_c}\Big[\Ln{\frac{64 x_c^4}{1+y_c}}+8 x_c^2 \Ln{\frac{1+y_c}{64 x_c^4}}\Big]+9 \Ln{\frac{-1+2 x_c^2+y_c}{x_c^2}}\\
-10 y_c \Ln{\frac{x_c m_b}{\mu}}+12 x_c^2 y_c \Ln{\frac{x_c m_b}{\mu}}+4 x_c^2 \Ln{\frac{1+y_c}{2 x_c}} \\
\leri{31+12 x_c^2 \Ln{\frac{x_c m_b}{\mu}}}+4 \Ln{\frac{2 x_c}{1+y_c}}\Big[-7+39 x_c^4+6 \leri{2 x_c^2-1} \Ln{\frac{x_cm_b}{\mu}}\Big]\bigg]\\
+24 x_c^2 \leri{1-8 x_c^2}\Li{\frac{1}{2}\leri{1-y_c}}\Bigg\}
.\label{eq:gc226}\end{array}\end{equation}

%% References


%%\bibliographystyle{model1a-num-names}
%%\bibliography{sources}




  %bibtextest

\end{document}